\documentstyle[preprint,aps]{revtex}
\begin{document}
\draft
\title{Study of Pairing Correlations in the Attractive Hubbard Model on
Chains, Ladders, and Squares}
\author{M. Guerrero, G. Ortiz, and J. E. Gubernatis}
\address{Theoretical Division, Los Alamos National Laboratory, Los
Alamos, NM 87444}
\date{\today}
\maketitle
\begin{abstract}
We report the results of zero temperature quantum Monte Carlo
simulations and zero temperature mean-field calculations of the
attractive Hubbard model on chains, ladders, and square lattices. We
investigated the predictability of the BCS approximation, the
dimensional cross-over of the pairing correlation function from one to
two dimensions as a function of the ladder width, and the scaling of
these correlations to the thermodynamic limit of the two-dimensional
model. We found that the BCS wave function is quantitatively correct
only for small values of $U/t$. For the system sizes, electron
fillings, and interaction strengths studied, we never saw the
dimensional cross-over. In general our ability to achieve the
dimensional cross-over and accurate scaling to the thermodynamic limit
was limited by the size of the systems we could simulate. For these
sizes, although we saw the necessary signature of ODLRO, the properties
of the model did not vary monotonically with increasing system size
because of shell effects.  We contrast this situation with the
dimensional cross-over and scaling to the thermodynamic limit of the
Ising model.
\end{abstract}

\section{Introduction}
We report the results of zero temperature quantum Monte Carlo (QMC)
simulations and zero temperature mean-field calculations of the
attractive Hubbard model on chains, ladders, and square lattices. We
investigated the predictability of the BCS approximation, the
dimensional cross-over of pairing correlations from one to two
dimensions as a function of the ladder width, and the scaling of these
correlations to the thermodynamic limit of the two-dimensional model.
We found that the BCS wave function is quantitatively correct only for
weak attractive interactions. For the system sizes, electron fillings,
and interaction strengths studied, we never saw the dimensional
cross-over, and we were unable to scale the pairing correlations on
finite-size square lattices to those on infinitely-sized ones. In
general our ability to achieve the dimensional cross-over and the
proper scaling to the thermodynamic limit was limited by the size of
the systems we could simulate. For these sizes, although we saw the
necessary signature of off-diagonal long-range order (ODLRO), the
properties of the model did not vary monotonically with increasing
system size because of shell effects.

The BCS wave function occupies a deservedly prominent place in the
theory of superconductivity. For traditional superconductors, the
electron-phonon interaction induces an attractive interaction between
pairs of electrons that is well described by the reduced BCS
Hamiltonian \cite{schrieffer}
\begin{equation}
 H = \sum_{{\bf k}\sigma} \epsilon_{\bf k}n_{{\bf k},\sigma} 
     -\sum_{{\bf k},{\bf k}'}V^{}_{{\bf k},{\bf k}'}
     c^\dagger_{{\bf k},\uparrow}c^\dagger_{-{\bf k},\downarrow}
     c^{}_{-{\bf k}',\downarrow}c^{}_{{\bf k}',\uparrow} \ ,
\label{eq:attr}
\end{equation}
and the BCS wave function
\begin{equation}
 |{\rm{BCS}}\rangle = \prod_{\bf k}(u^{}_{\bf k}+v^{}_{\bf k}
      c^\dagger_{{\bf k},\uparrow}c^\dagger_{-{\bf k},\downarrow})|0\rangle
\label{eq:bcswfn}
\end{equation}
provides an excellent approximation to the eigenstates of this
Hamiltonian. In fact, under a number of circumstances, it is the exact
solution in the thermodynamic limit \cite{bursill91}.

Characteristic of BCS-like superconductors is the large number of
Cooper pairs inside the volume defined by the coherence length $a_0$
of a pair. In high temperature (cuprate) superconductors, as well as
several other classes of much lower temperature superconductors,
including doped cuprate ladders \cite{uehara96}, the coherence length is
comparatively small. For these materials, one often proposes the
attractive Hubbard model
\begin{equation}
  H =-t\sum_{\langle ij \rangle,\sigma} (c_{i,\sigma}^\dagger c_{j,\sigma}
    + c_{j,\sigma}^\dagger c_{i,\sigma})
    - U \sum_{i} n_{i,\uparrow} n_{i,\downarrow} \ ,
\label{eq:hubbard}
\end{equation}
where $U>0$, as a counterpart to the BCS reduced Hamiltonian.  Indeed,
for any allowable electron filling, two electrons are energetically
favored to pair onto a lattice site, thus giving them a seemingly very
short coherence length. For dimensions of two and higher, the
ground-state possesses s-wave ODLRO in the thermodynamic limit over
certain ranges of electron fillings if the lattice is bipartite and the
number of $A$ sites does not equal the number of $B$ sites, a condition
not satisfied, for example, by a square lattice \cite{shen93}. 

In this work we studied the attractive Hubbard model on chains, ladders
and squares, using an exact QMC method. The attractive Hubbard model
has no fermion sign problem so it is relatively easy to study by QMC
simulations. One goal was to observe the dimensional cross-over of the
superconducting pairing correlation functions from the well-known
power-law behavior in one dimension to ODLRO in two dimensions. We
studied mostly the quarter-filled system and compared our results with
the predictions of the BCS approximation. We tried to establish the
range of validity of this approximation, and for the square lattices we
tried to investigate both the weak and strong coupling regimes to
assess the validity of the argument of Nozi\`eres and Schmitt-Rink
\cite{nozieres85} regarding the smooth cross-over from one regime to
the other. 

Nozi\`eres and Schmitt-Rink \cite{nozieres85} studied the transition
from weak to strong coupling superconductivity in an attractive
continuum fermion model
\begin{equation}
H = \sum_{{\bf k},\sigma}\frac{k^2}{2m} n_{{\bf k},\sigma}
          - \sum_{{\bf k},{\bf k}',{\bf q}} V_{{\bf k},{\bf k}'}^{}
c^\dagger_{{\bf k+q}/2,\uparrow}c^\dagger_{-{\bf k+q}/2,\downarrow}
c^{}_{-{\bf k'+q}/2,\downarrow}c^{}_{{\bf k'+q}/2,\uparrow} \ ,
\end{equation}
and the attractive lattice Hamiltonian (\ref{eq:hubbard}). For both
models, they argued that the BCS wave function is a good approximation
in the weak and strong coupling limits. In the continuum model, weak
coupling  corresponds to carrier densitites $\langle n \rangle$
satisfying $\langle n \rangle a_0 \gg 1$, and strong coupling
satisfying the opposite limit. For the lattice model, coupling strength
corresponds to the relative size of the interaction strength $U$ and
the non-interacting bandwidth $W$: Strong coupling is $U\gg W$ and weak
coupling is $U\ll W$. Further they argued that this approximation
interpolates smoothly and accurately between these limits, a remarkable
fact because of the distinctly different physics occurring at these
limits: For weak coupling, Cooper pairs form according to the usual BCS
picture; for strong coupling, (Schafroth) pairs are hard-core bosons
which may Bose condense.

We found the weak coupling regime for a quarter-filled system, i.e.
$\langle n\rangle = 1/2$, on a square lattice is limited to values of
$U/t$ satisfying $0\leq U/t \lesssim 1$. For larger values of $U/t$,
significant disagreements in the values of on-site s-wave pairing
correlations existed between the results of the simulations and the
predictions of the BCS approximation. These disagreements remained for
values of $U/t$ as large as 8, which is as large as we could push our
simulations,  and suggest the strong coupling regime, if the arguments
of Nozi\`eres and Schmitt-Rink are correct, lies well beyond $U/t=8$.
(With our numerical technique we can only accurately simulate values of
$U$ comparable to the size of $W$.) 

We comment that in the strong coupling limit ($U/t \gg 1$ and with any
band filling of an even number of fermions), the attractive Hubbard
model, up to second order in degenerate perturbation theory, maps to an
interacting hard-core boson system in the subspace of doubly occupied
and empty sites \cite{Micnas}. These bosons hop to nearest neighbors
with an effective amplitude $-2 t^2/U$ and interact with a nearest
neighbor repulsion $2 t^2/U$. However, the BCS wave function is the
exact eigenstate of this effective problem only in the completely empty
and completely filled lattice cases (see Appendix A). From our analysis
we are unable to say  whether away from the extreme cases the BCS fixed
point is the correct strong coupling limit. The BCS fixed point can still
be the correct one and large quantitative differences can exist. 

Our main object of study was pairing correlation functions.  We found
that the distance dependent pairing properties of the model on a square
lattice did not vary monotonically as we systematically expanded its
size, and we were unable to  obtain a very accurate estimate of the
order parameter in the thermodynamic limit, even though it is clear
from the calculations that the system has ODLRO in that limit. 
This shortcoming contrasts a much earlier zero temperature QMC study
\cite{scalettar89} that claimed to do such an extrapolation on the
integrated pairing correlation function and obtained a value for the
order parameter consistent with our results. Our computed results for
the long-range spatial dependence of the s-wave pairing correlations
for a long chain compared very favorably with the Bethe {\em ansatz}
predictions for a chain in the thermodynamic limit. Systematically
coupling these long chains did not produce a monotonic variation in
these correlations, and computational costs forced us to limit our
study to a $L\times 4$ systems with $L\sim 50$. We believe the
irregular variations in the properties of the rectangular and square
structures are consequences of shell effects characteristic of
finite-sized systems of fermions. We remark that increasing the value
of $U/t$ reduces the shell effects as one would expect.  These
finite-size shell effects prevented us from observing the cross-over to
the two dimensional behavior in the correlation functions. Wider
systems or larger values of the interactions are needed to  observe
such cross-over.

In the next section we summarize several useful exact properties of
the attractive Hubbard model in two-dimensions, and then in
Section~III, we summarize our numerical approach. In Section~IV we
present our main results, and in the final section, Section V, we will
discuss the implications of our results and the effects of finite-size
on other simulations of interacting electron systems.

\section{Attractive Hubbard Model}

A number of exact statements can be made about the properties of the
attractive Hubbard model at zero temperature. In one dimension the
most important statements for present purposes are the absence of any
LRO \cite{liebwu} and the inverse power-law form for the pairing
correlation function at large distances between pairs
\cite{kawakami91}. In this Section we will focus on those statements
that are at least true for ground state properties in two dimensions.
Most are in fact true for any dimension. Most assume a bipartite
lattice. Unfortunately, we are unaware of any exact results assuring
the model on a square lattice has ODLRO. The Mermin-Wagner theorem
\cite{mermin} only assures that LRO cannot exist in two-dimensions at
finite temperature. We are also unaware of any pertinent theorems
regarding the attractive Hubbard model on ladder structures.

Perhaps the most fundamental theorem says the ground state for an even
number of electrons is a non-degenerate singlet for every electron
filling \cite{lieb89}.  On the other hand, the most useful property of
the model is its mapping for any electron filling onto a half-filled
repulsive Hubbard model in a magnetic field \cite{auerbach}. Charge
and pairing operators map into pseudo-spin operators and vice versa.
A specific version of this mapping maps staggered components of the
pseudo-spin operators in the half-filled repulsive model into charge
and pairing operators in the attractive half-filled model
\cite{singh91}.

At half-filling, on-site s-wave ODLRO exists in the attractive Hubbard
model if and only if a charge-density wave (CDW) LRO exists
\cite{shen96}.  It is generally accepted that the ground state of the
half-filled, two-dimensional, repulsive model is one of
anti-ferromagnetic (AF) LRO. Because longitudinal pseudo-spin maps to
charge, and transverse pseudo-spin, to pairing, the acceptance of AF
LRO for the half-filled repulsive model seems sufficient to establish
on-site s-wave ODLRO and CDW LRO as ground states of the half-filled
attractive model.  Indeed numerical studies support this conclusion
\cite{scalettar89} .

Another theorem states if the charge spectrum is gapped in either the
repulsive or attractive models, then there can be no on-site s-wave
ODLRO \cite{momoi95}. It is generally accepted that the ground state
of the half-filled, two-dimensional, repulsive model has a charge
gap. This presence implies a spin gap for the ground state of the
half-filled, two-dimensional, attractive model. Thus the theorem does
not violate the inference of ODLRO in the attractive model at
half-filling.

In this paper we are exclusively interested in the properties of the
attractive model away from half-filling. Here a theorem states that
on-site s-wave ODLRO exists if and only if extended s-wave ODLRO exists
\cite{zhang90}. Noteworthy is that this theorem also states that the
magnitude of the order parameter for the extended s-wave state vanishes
as half-filling is approached, making the doped case essentially
different in character from the half-filled case. In particular, if
$\langle \Delta_s\rangle$ is the expectation value of the on-site
s-wave order parameter and $\langle \Delta_{s^*}\rangle$ is that of
extended s-wave, then
\begin{equation}
 \langle \Delta_{s^*}\rangle = \frac{-U-2\mu}{t}\langle \Delta_s\rangle
\label{eq:zhang}
\end{equation}
where $\mu$ is the chemical potential. At half-filling, because of
particle-hole symmetry, $2\mu = -U$. Accordingly, at half-filling
$\langle \Delta_{s^*}\rangle = 0$. We comment that this relation
between  the extended s-wave order parameter and the on-site
(isotropic) s-wave order parameters depends on the hopping integral
being isotropic. If one were to set, for example, $t_x=t$ and $t_y=-t$,
the d$_{x^2-y^2}$-wave order parameter would appear on the left hand
side of (\ref{eq:zhang}). In the isotropic hopping case we did not see
any significant signal of d-wave pairing. In Appendix~A, we prove the
equivalent theorem for the BCS approximation.

Ground-state numerical studies away from half-filling see a necessary
signature of on-site s-wave ODLRO order \cite{scalettar89}. The
pair-pair correlation function clearly does not vanish when the pairs
are well separated. To establish ODLRO one also needs to show that
this correlation also does not vanish in the thermodynamic
limit. Establishing this for the two-dimensional attractive model was
one of our goals. To achieve this objective we used an exact
QMC method, that is, a method with no sign problem. We
now summarize this method.

\section{Numerical Approach}

The details of our numerical approach are the same as those for the
constrained-path Monte Carlo (CPMC) method \cite{zhang95} except we have
no constraint 
because the simulations have no sign problem. Because of this, apart
from statistical error, the method is exact. Briefly the strategy of
our approach is as follows:

Starting with some trial state $|\psi_T\rangle$, we project out the
ground state by iterating
\begin{equation}
 |\psi'\rangle = e^{-\Delta\tau (H-E_T)}|\psi\rangle \ ,
\end{equation}
where $E_T$ is some guess of the ground-state energy.  Purposely
$\Delta\tau$ is a small parameter so for $H=T+V$ we can write
\begin{equation}
 e^{-\Delta\tau H}\approx e^{-\Delta\tau T/2}
	                  e^{-\Delta\tau V}
                          e^{-\Delta\tau T/2} \ ,
\end{equation}
where $T$ and $V$ are the kinetic and potential energies. We used values
of $\Delta\tau$ ranging from 0.03 to 0.05.

For the study at hand, the initial state $|\psi_T\rangle$ is the direct
product of two spin Slater determinants, i.e.,
\begin{equation}
 |\psi_T\rangle = \prod_\sigma |\phi_T^\sigma\rangle \ .
\end{equation}
Because the kinetic energy is a quadratic form in the creation and
destruction operators for each spin, the action of its exponential on
the trial state is simply to transform one direct product of Slater
determinants into another. While the potential energy is not a
quadratic form in the creation and destruction operators, its
exponential is replaced  by sum of exponentials of such forms via
the discrete Hubbard-Stratonovich transformation
\begin{equation}
 e^{\Delta\tau U n_{i,\sigma}n_{i,-\sigma}} 
=\frac{1}{2}e^{-\frac{1}{2}\Delta\tau U}
 \sum_{x=\pm 1}e^{-x         \Delta\tau J(n_{i,\sigma}+n_{i,-\sigma}-1)}
               e^{\frac{1}{2}\Delta\tau U(n_{i,\sigma}+n_{i,-\sigma})}
\end{equation}
provided $U\ge 0$ and $\cosh \Delta\tau J = e^{\Delta\tau
U/2}$. Accordingly we re-express the iteration step as
\begin{equation}
 \prod_\sigma |\phi_\sigma'\rangle = \int d\vec x\, P(\vec x)
    \prod_\sigma B_\sigma(\vec x)|\phi_\sigma\rangle \ ,
\end{equation}
where $\vec x =(x_1,x_2,\dots,x_N)$ is the set of Hubbard-Stratonovich
fields (one for each lattice site), $N$ is the number of lattice
sites, $P(\vec x)=(\frac{1}{2})^N$ is the probability distribution for
these fields, and $B_\sigma(\vec x)$ is an operator function of these
fields formed from the product of the exponentials of the kinetic and
potential energies.

Upon examination, one sees that $B_\sigma(\vec x)=B_{-\sigma}(\vec x)
\equiv B(\vec x)$. This equivalence means the equivalence of the
propagators for the separate spin components of the initial state. If
these components are identical, i.e., if $|\phi_T^\sigma\rangle
=|\phi_T^{-\sigma}\rangle \equiv |\phi_T\rangle$, then only one
component needs to be propagated
\begin{equation}
 |\phi'\rangle = \int d\vec x\, P(\vec x)B(\vec x)|\phi\rangle
\end{equation}
but more importantly, the overlap of the initial state with the current
state $\langle \psi'|\psi_T\rangle = \langle\phi'|\phi_T\rangle^2$ and
thus is always positive. This positivity is sufficient to eliminate the
sign problem.

The Monte Carlo method is used to perform the multi-dimensional
integration over the Hubbard-Stratonovich fields. It does so by
generating a set of random walkers initialized by replicating
$|\psi_T\rangle$ many times. Each walker is then propagated independently by
sampling a $\vec x$ from $P(\vec x)$ and propagating it
with $B(\vec x)$. After the propagation has ``equilibrated,'' the sum
over the walkers provides an estimate of the ground-state wave function 
$|\psi_0\rangle$.

In practice we performed an importance-sampled random walk, obtained by
defining for each Slater determinant $|\phi\rangle$ another one
$|\tilde\phi\rangle$ via
\begin{equation}
 |\tilde\phi\rangle = \langle\phi_T|\phi\rangle|\phi\rangle \ ,
\end{equation}
and using the transformed iterative equation
\begin{equation}
 |\tilde \phi'\rangle = \int d\vec x\, \tilde P(\vec x)B(\vec x)|
                \tilde\phi\rangle \ .
\end{equation}
In this equation
\begin{equation}
\tilde P(\vec x) = P(\vec x)
                 \frac{\langle\phi_T|\phi'\rangle}
                      {\langle\phi_T|\phi\rangle} \ .
\end{equation}
Thus importance sampling changes the probability distribution of the
Hubbard-Stratonovich fields, biasing it towards the generation of
states with large overlap with the initial state.  The branching
nature of the random walk is the same as described for the CPMC method
and will not be discussed here.  It is a necessary procedure for
controlling the variance of the computed results.

We used two different estimators for the expectation values of some
observable ${\cal O}$. One is the mixed estimator
\begin{equation}
 \langle {\cal O}\rangle_{\mathrm{mixed}} =
     \frac{\langle\psi_T|{\cal O}|\psi_0\rangle}
          {\langle\psi_T|\psi_0\rangle} \ ,
\end{equation}
and the other is the back-propagated estimator
\begin{equation}
 \langle{\cal O}\rangle_{\mathrm{bp}} =
     \frac{\langle\psi_T|e^{-\ell\Delta\tau H}{\cal O}|\psi_0\rangle}
          {\langle\psi_T|e^{-\ell\Delta\tau H}|\psi_0\rangle} \ ,
\end{equation}
where $|\psi_0\rangle$ is the QMC estimate of the ground state and
$\ell$ is typically in the range of 20 to 40.  For observables that
commute with the Hamiltonian, the mixed estimator is a very accurate
one and converges to the exact answer as $|\psi_0\rangle$ converges to
the exact ground state. For observables that do
not commute with the Hamiltonian, like correlation functions,
the back-propagated estimator has been found to give very accurate
estimates of ground-state properties. Significant differences between the
predictions of these two estimators often exist.

We remark that we could have projected to the ground state using the
BCS wave function as our starting point \cite{guerrero99}. This wave
function is not normally represented as a direct product of two spin Slater
determinants, but by a trick proposed by Yokayama and Shiba
\cite{yokoyama88}, one can re-express it as a single Slater
determinant. In this new representation, one can then show that there
is still no sign problem, but that the computational cost of working
with this wave function is at least a factor of 4 more. Since no
statistically significant differences in computed results occur, we
almost exclusively used the computationally more efficient direct
product of free-electron wave functions for
$|\psi_T\rangle$. Additionally, most of our calculations were done for
closed-shell electrons fillings, i.e., non-degenerate electron
fillings in the non-interacting problem. For these fillings and for
the same amount of computing time, the statistical error of our
expectation values is considerably smaller than that of our
expectation values at open-shell fillings (which can be easily handled
when the BCS wave function is used as constraint \cite{guerrero99}).

We also remark that the standard auxiliary-field projector Monte Carlo
method also has no sign problem for the attractive Hubbard
model. Although only a few comparisons exist, we are unaware of any
statistically significant differences between results from that method
and results from the one used here. If both are exact procedures,
then, of course, there should not be any. The estimators used in the
two methods however are different as the auxiliary-field projector
method uses a different back-propagation estimator for all
observables.

\section{Results}

In sequence we will now report and discuss results for the attractive
Hubbard model on chains, squares, and ladders. In each case we will
comment on the predictions of the BCS approximation relative to the
predictions of the QMC simulations. For chains we will also comment on
supportive calculations we performed using the density matrix
renormalization group (DMRG) method \cite{dmrg}. Included in a
separate subsection is a discussion on the specific issue of
dimensional cross-over.

Because the variances of our computed results are smaller for
closed-shell fillings, we only considered such fillings, and because the
shell structure changes with lattice size, as we changed lattice
sizes, we could not maintain the electron densities $\langle n
\rangle$ at a fixed value. For the results reported here, we fixed the
electron density as close as possible to the quarter-filled value, 
that is, to a value of 1 electron per 2 lattice sites.  
The actual fillings for many of the 
lattices studied are given in Table~1. In all cases we had
equal numbers of up and down spin electrons.
In one-dimension we used $\langle n \rangle = 1/2 + 1/L$
which converges to quarter filling as $L$ is increased.
We remark that the accuracy of our QMC calculations were benchmarked
against those of exact diagonalizations calculations of $4\times 4$
lattices. The QMC results, and in particular those computed by
the back-propagation estimators, agreed well within statistical error with
those of the more precise method. 

\subsection{Chains}

For both finite and infinite chain lengths one can obtain the energies
of the attractive Hubbard model from its Bethe {\em ansatz} solution
\cite{liebwu}. Marsiglio and Tanaka \cite{marsiglio97,tanaka99}, for
example, have made extensive comparisons of these energies with the
predictions of the BCS approximation. Additionally, by computing the
Bethe {\em ansatz} energies for systems of $N_e$, $N_e-1$ and $N_e+1$
electrons, they also calculated the pair binding energy by evaluating
$\Delta_{N_e} = (E_{N_e-1}-2E_{N_e}+E_{N_e+1})/2$, which in the BCS
approximation is the gap $\Delta$ \cite{lee88}. The energy is the
least accurate at intermediate coupling strengths at half-filling,
where the disagreement however is only a few per cent, and becomes
exact in the weak and strong coupling limits at low electron
density. In general, the pair binding energy is poorly reproduced,
with the disagreement being the least in the dilute electron density
limit. Typically, the BCS wave function overestimates the binding
energy. For finite-size systems they also found significant shell
effects in the energy as a function of electron density or as a
function of chain length in the weak coupling regime. They did not
investigate the pairing correlations.

Exact analysis of the Bethe {\em ansatz} solution by Kawakami and Yang 
\cite{kawakami91} showed that in the thermodynamic limit the on-site
s-wave pairing correlation function $P_s(R)$ behaves like
\begin{equation}
  P_s(R) = \langle \Delta_s^\dagger(R)\Delta_s(0)\rangle 
      \sim \frac{1}{R^\beta}
\label{eq:power}
\end{equation}
at large separations $R$ between the pairs.  The non-universal
exponent $\beta$ is a function of both $U/t$ and the electron
filling. (Kawakami and Yang provide a graph giving $\beta$ for a
selection of values of $U/t$ for all fillings between 0 and 1.) Because
the BCS approximation always predicts ODLRO, its predictions for the
behavior of pairing correlation functions is qualitatively incorrect.

We computed the on-site s-wave pairing function in several different
ways. In the QMC simulations, we used periodic boundary conditions,
took
\begin{equation}
   \Delta_s(i) = c_{i,\downarrow}c_{i,\uparrow} \ ,
\end{equation}
and then calculated
\begin{equation}
  P_s(R) = \frac{1}{L}\sum_i \langle
                  \Delta_s^\dagger(i+R)\Delta_s(i)\rangle
\label{eq:standard}
\end{equation}
for each possible value of $R$ by using the back-propagation
estimator. In the DMRG calculation, we used open boundary conditions
and computed $P_s(R)$ in two different ways. In the first way we
computed
\begin{equation}
  P_s(R) = \langle\Delta_s^\dagger(i_R+R)\Delta_s(i_R)\rangle
\label{eq:center}
\end{equation}
for each value of $R$ relative to some site $i_R$, which depends on
$R$, chosen to place the pairs as close to the center of the chain as
possible. In other words, to avoid edge effects, for a given value of
$R$, we used the sites closest to the center. At the large values of
$R$, the resulting estimate displayed rapid fluctuations that made
comparisons with the QMC and Bethe {\em ansatz} predictions
difficult. We found we could smooth out the fluctuations by computing
$P_s(R)$ from \cite{white}
\begin{equation}
 P_s(R) = \frac{1}{N_p(R)}\sum_{i=1}^{N_p(R)} 
         \langle\Delta_s^\dagger(R)\Delta_s(0)\rangle_i \ .
\label{eq:average}
\end{equation}
Here we sum over all possible pairs of lattice sites separated by the
same displacement $R$ and divide this sum by the number $N_p(R)$ of such
pairs of sites, which is a function of $R$. At large values of $R$,
this function was much smoother. More importantly, it compared very
well with the QMC results.

In Figs.~1 and 2 we show samples of our results for $U/t=2$. In Fig.~1,
we superimpose several calculations of $P_s(R)$ for different chain
lengths $L$ to illustrate that increasing the chain length did not
change the pairing correlations at short and intermediate distances. It
simply increased the range in distance over which we correctly captured
these correlations. We fitted the large distance behavior of the
correlation functions to the inverse power-law function
(\ref{eq:power}) and found exponents $\beta$ consistent with the Bethe
{\em ansatz} prediction. For example, from Kawakami and Yang's graph,
we estimated $\beta = 0.80(2)$. Our fit for the chain of $L=66$, shown
in the inset to Fig.~1, predicted $\beta=0.79(3)$. Exponents for other
fits are given in Table~1.  For long enough chains we obtained close
agreement between the value of $\beta$ obtained by fitting and the
value obtained from Kawakami and Yang's graph.  In performing the fit
we eliminated the short-range behavior and the up-turn in the
correlations at large distances. This up-turn is a finite-size effect
caused by periodic boundary conditions.

A comparison of our QMC results with those from BCS and DMRG
calculations are shown in Fig.~2a. Clearly the predictions from the
two quite different numerical methods agree, provided we average the
DMRG results as discussed above. For purposes of comparison, we also
show the prediction of the BCS approximation. It is qualitatively
different from what is required from the exact
solution. Fig.~2b shows the difference in the DMRG results when
computed for the two different ways of averaging, that is, when computed
from (\ref{eq:center}) and (\ref{eq:average}).

In short, both the DMRG and QMC predictions agree well with each other
and with the exact prediction of the Bethe {\em ansatz} solution for
the asymptotic form of the pairing correlation function at large
distances. As expected, the BCS approximation is qualitatively
incorrect for this same function.

\subsection{Squares}

In two dimensions, an exact analytic solution for the Hubbard
model does not exist. Within statistical error, the QMC approach
however provides an exact numerical solution. In this subsection we
are principally concerned with comparing predictions of the BCS
approximation with those of the exact QMC simulations.

In Fig.~3, we show the QMC and BCS $P_s(R)$ as a function of the
distance $R$ for an $8\times 8$ lattice with $U/t=1$, $2$, and $4$. The
curves show the same behavior: The correlations rapidly drop in
magnitude over a distance of a few lattice spacings and then stay
constant over the remainder of the distances. The constant behavior at
these larger distances is a signature of ODLRO, if the distances
extended to infinity. As the results go from $U/t=1$ to $U/t=4$, the
agreement between the QMC simulations and the BCS approximation for
lattices of the same size progressively worsens.  Beyond $U/t=1$, the
quantitative agreement is poor.  One also sees that as $U/t$ increases,
the magnitude of the pairing correlations increases. At large
distances, if there is ODLRO, the pairing correlation function becomes
asymptotically equal to $\langle\Delta_s\rangle^2$. In the BCS
approximation, $U$ times $\langle\Delta_s\rangle$ equals the energy
gap, and this gap becomes exponentially small when decreasing $U/t$ to 0
and proportional to $U/t$ when increasing $U/t$ to infinity.  However, as
noted before, the attractive Hubbard model is gapless if it has
on-site s-wave ODLRO \cite{momoi95}.
 
While we computed several cases for $U/t=8$, the statistical error was
considerably larger and hence we do not show these results. However in
all cases computed, as we increased the magnitude of $U/t$, we never saw
the QMC and BCS results approach each other. In fact, as we
increased $U/t$ beyond 1, their difference continuously increased. 
In Fig.~4, we show both the on-site and extended s-wave pairing
correlation functions as a function of distance for $U/t=4$ and a $14\times
14$ lattice.  In the inset we eliminated the short distance part
to emphasize
the flat large distance behavior for both correlation functions. By
averaging over this flat behavior we estimated both $\langle
\Delta_{s^*}\rangle^2$ and $\langle \Delta_s\rangle^2$ and attempted
to verify that the ratio is proportional to $({-U-2\mu}/{t})^2$ 
as stated by Eq. (\ref{eq:zhang}).  
Because most of our simulations were done for fixed particle numbers 
and closed-shell fillings, the estimation of $\mu$ was difficult. 
Using a variation of the present method  \cite{guerrero99} that allows
an estimation of $\mu$, we found $\mu= -2.77$ for a $8\times 8$ system 
with $U/t=4$. Using this same value of $\mu$ for the $14\times 14$
system with $U/t=4$ yields $({-U-2\mu}/{t})^2= 2.37$ which compares
very well to the value of 2.40(3) estimated from the figures.  We also
observed that the ratio of the expectation values for the on-site and
extended s-wave order parameters was {\em exactly} obeyed in the BCS
approximation, provided the BCS value of $\mu$ was used. In Appendix~A 
we present the analysis leading to this result. 
We also observed that the long range values of both the on-site and
extended s-wave pairing correlation functions do not vary
monotonically with lattice size. We illustrate this in Fig.~5 by
plotting both the QMC and BCS results for $\langle\Delta_s\rangle$
as a function of the reciprocal of the number $N$ of lattice sites for
both $U/t=2$ and $U/t=4$. Within the BCS approximation one can compute
$\langle\Delta_s\rangle$ directly for any lattice size.

In Fig.~5a, the BCS result for $U/t=2$ clearly does not converge to the
thermodynamic limit monotonically. The same lack of monotonicity is
apparent in the variation of the QMC results. Because the BCS and QMC
results are quantitatively different, we see no reason why the QMC
results might extrapolate to the BCS result, but more importantly we
do not see how to extract an accurate value of
$\langle\Delta_s\rangle$  in the thermodynamic limit, although it is
clear from our results that  it is a finite value and therefore there
will be ODLRO in the thermodynamic limit.
In Fig.~5b, we show similar curves for $U/t=4$. The apparent
monotonicity of the BCS result is deceptive, as illustrated in the
inset to this figure. The more important point is that the BCS results
approach the thermodynamic limit for $U/t=4$ case more rapidly than
they did for the $U/t=2$ case. This trend is very general and suggests
that the QMC results for larger values of $U/t$ will also converge more
rapidly to the thermodynamic limit. Large interactions reduces fermion shell
effects. Unfortunately both the necessarily large values of $U/t$ and
lattice sizes appear inaccessible with our simulation method. For
$U/t=1$ we estimate that the BCS approximation has approached the
thermodynamic limit within a per cent for lattices of the order of
$60\times 60$; for $U/t=8$ our estimate is for lattices of the order
of $10\times 10$. As mentioned before, at $U/t=8$ the QMC results have
large statistical errors in the back-propagation estimate of the
pairing correlation functions.

In summary, for on-site s-wave pairing correlations, we find that the
QMC and BCS results agree in the weak coupling limit. While the QMC
results show a necessary signature of ODLRO, we were unable to
extrapolate accurately our results for the order parameter to the
thermodynamic limit.

\subsection{Ladders}

An alternate way to reach the two-dimensional thermodynamic limit is
extrapolating from a ladder structure of fixed, but long, length $L$ by
letting the number of legs $M$ become large. Here we report our results
from the use of this strategy.

We built up the ladder structures by coupling 1, 2, and 3 additional
chains to the one-dimensional chain and making the interchain and
intrachain hopping amplitudes equal. We used periodic boundary
conditions in the $x$ (long) direction and open boundary conditions in
the $y$ (short) direction.  We call this a cylindrical boundary
condition. For the on-site s-wave pairing correlation function we
computed it only along the $x$-direction, using
\begin{equation}
  P_s(R) = \frac{1}{L}\sum_{i_0} \langle
                  \Delta_s^\dagger(i_0+R)\Delta_s(i_0)\rangle \ ,
\end{equation}
where the $i_0$ are lattice sites along one of the central legs of the
ladder. For a 3-legged ladder, the sites in the $x$-direction are on the
central leg. For a 2-legged ladder, we used just one of the legs. By
symmetry, the chain-directed pairing correlations are the same on
each leg. For a 4-legged ladder, we just used one of the two central
legs.

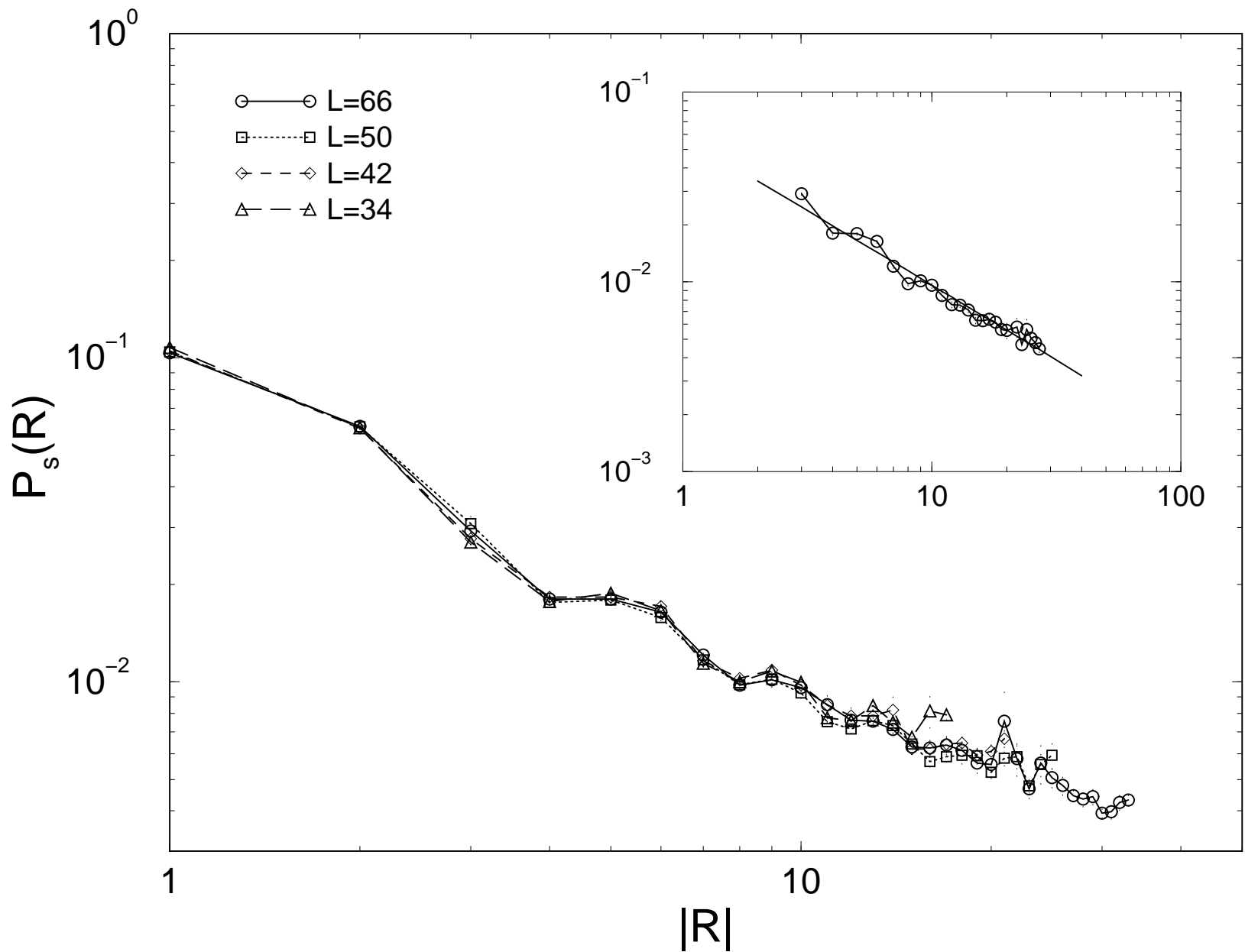
\begin{figure}
\caption{The QMC on-site s-wave pairing correlation function $P_s(R)$ as a
function of the distance $|R|$ between pairs for a $U/t=2$ chain of
different lengths $L$. The inset shows the fit of results for the $L=66$
chain to the inverse power law form (\ref{eq:power}) using a value
of $\beta=0.79$.}
\label{fig1}
\end{figure}

To provide some perspective, we first present in Fig.~6 plots of the
energy as a function of system size. In particular, in Fig.~6a is the
energy per site for a $U/t=2$ chain as a function of the reciprocal of
chain length. The variation with length is smooth and monotonic.
Linearly extrapolating this data to infinite length, we find an energy
per site of $-1.079(1)$ which is within statistical error of the Bethe
{\em ansatz} value of -1.079 \cite{marsp}. In Fig.~6b is the energy per
site as a function of the reciprocal of the number of legs $M$ for a
ladder of length $L=50$. The variation is smooth and monotonic and
extrapolates to a value of -1.49(1), a value similar to the one we
found from the QMC simulations for the two-dimensional lattice. This is
not surprising since, although the extrapolating function can be
shape-dependent, the energy is a bulk quantity, i.e., independent of
the way one performs the thermodynamic limit.  These two-dimensional
QMC results are shown in Fig.~6c for the non-interacting case and for
the interacting case computed by both the BCS approximation and the QMC
method. The variation is neither smooth or monotonic. Additionally it
tracks the non-interacting problem. This tracking supports a claim that
the non-monotonic variations are caused by shell effects that are
obvious and inherent to non-interacting fermion problems.
For the s-wave pairing correlation function, a typical result is shown
in Fig.~7. Similar to the one-dimensional case (Fig.~\ref{fig1}), the
curves for increasing ladder length lie on top of those for shorter
lengths. In contrast to the one-dimensional case, however, distinct
oscillations exist but the overall trend is a decrease in magnitude
with increasing distance.  For the 2 and 3-legged cases we fitted the
top and bottom envelopes of the oscillations to the power law decay
(\ref{eq:power}) and found the two decays rates to be close but not
the same. These exponents were also close to the one-dimensional
value (Table 1). An important observation is the overall magnitude of the
pairing correlations at first decreasing below the one-dimensional
value and then, as we built-up the ladder structure to 4-legs,
increasing above the 3-leg value. For a ladder of length $L=50$, the
variation of the s-wave pairing correlation function with the number
of legs is shown in Fig.~8a.  A typical fit to determine the exponents
of decay is shown in 
the inset. Other fitted values are given in Table~1.
We note that while the exponent does not change much from the one-dimensional
case to the 2 and 3 legged ladder cases, it does decrease
significantly when the 4th leg is added. 

We remark that using the BCS approximation, we computed the pairing
correlations for the same system sizes but with periodic boundary
conditions in both directions. While this approximation always showed
ODLRO, its order parameter $\langle\Delta_s\rangle$ showed the same
trend as we obtained from the QMC simulations: In going from 1 to
3-legs, the value of the order parameter at first monotonically
decreased but then increased relative to the value of the 3-legged
ladder when the fourth leg was added.

\subsection{Dimensional Cross-Over}

What type of variation with ladder width should one expect? Recently
finite temperature QMC studies of quantum spin ladders found
non-monotonic variations with the number of legs
\cite{frischmuth96,greven96,ammon99}. More specifically, the dynamic
susceptibilities of even-legged ladders had spin gaps while those of
odd-legged ones did not. The size of the spin gap rapidly decreased
with increasing number of legs and is a result of topological
considerations similar to those associated with Haldane gaps in
one-dimensional chains. For fixed, but large, ladder lengths the
uniform susceptibility vanished as the temperature was lowered and
then the number of legs extrapolated to infinity. This susceptibility
however extrapolated to a non-zero value if at low temperatures the
two-dimensional results were extrapolated to the thermodynamic limit.
Proper extrapolation of this system to the thermodynamic limit appears
to be an unsolved problem. The theoretical and experimental situation
has received several recent reviews \cite{dagotto96,rice98}.
 
Turning from quantum critical phenomena to the finite temperature
critical behavior and using the two-dimensional Ising model as an
example, we can easily study the dimensional cross-over from one to
two dimensions and the extrapolation of two-dimensional results to the
thermodynamic limit. We in fact were able to derive exact results,
given in Appendix~B, for the spin-spin correlations for such finite
systems. Using these exact results we computed the spin-spin
correlation function as a function of distance along the central leg
for several illustrative cases. For the calculations presented in
Fig.~9, we fixed the reciprocal of the critical temperature at its
bulk value of $\frac{1}{2}\ln(\sqrt{2}+1)\approx 0.44$, chose a long ladder
length of 200 lattice sites, and computed the correlation function for
various number of legs. What is seen is a smooth but gradual
transition from the expected exponential decay in one dimension to
the expected power-law decay in two dimensions. Proper extrapolation
of this system to the thermodynamic limit is an exactly solvable
problem \cite{mccoy}.

What is the relevance of the studies on these spins systems to our QMC
simulations?  From studies in finite-temperature critical phenomena,
one suggestion on the controlling physics of a dimensional cross-over
is the type of behavior depending on the relative size of the dominant
correlation length to the width of the system \cite{brankov}. This insight was
developed in part from analytic studies of $D$-dimensional ladders
sized $\infty\times\infty\times\cdots\times\infty\times M$ and
scaling its behavior to a $(D+1)$-dimensional systems by letting
$M\rightarrow\infty$. In going from one to two dimensions, as long
as one is on the critical surface, one expects the long range
behavior to be one-dimensional-like if the correlation length is
larger than the width and two-dimensional-like if it is smaller. The
various correlation functions we computed for the Hubbard model, such
as charge-charge, spin-spin, and pair-pair, rapidly decayed over a
distance of several lattice spacings from their on-site value,
suggesting our system sizes were large enough so that
if we were on the quantum critical surface, our physics would be more
two-dimensional-like than one-dimensional. We find that it appears to
be the other way around.

The picture so appropriate for finite-temperature critical phenomena
established for classical systems might not be appropriate to our
fermion simulations because the systems sizes are too small and are
dominated by shell effects, a phenomenon intrinsic to quantum fermion
systems. In Fig. 10, we plot the band-structure of the free-electron
case for 1, 2, 3, and 4 legged ladders of infinite length with
cylindrical boundary conditions. The horizontal line represents the
Fermi energy for an electron density of 1/2. For 1 leg, the
one-dimensional case, there is one-band. For 2-legs there are two
bands but only the lower band is filled. While the entire system is
quarter-filled, the one-dimensional-like lower band is
half-filled. Not surprisingly then is the one-dimensional-like decay
rate of the pairing correlations. For 3 and 4 legs there are 3 and 4
bands. More than one band is participating but the situation is not
yet two-dimensional-like. This simple band picture therefore provides some
qualitative basis for understanding our data. This understanding
suggests, at least for the sizes of the systems we were able to study,
that the length scale interpretation of dimensional cross-over is not
yet applicable.

As still another way of looking at the finite-size effects we present
Fig.~11 which displays the allowable {\bf k}-values for
non-interacting $2\times 8$, $4\times 8$, and $8\times 8$ lattices
with periodic boundary conditions. For quarter-filling the black dots
are occupied by an up and down spin electron, the gray dots are the
degenerate states in the open shell, and  the open dots are unoccupied.
Clearly for $2\times 8$ the occupied k-states are those of a
one-dimensional lattice. This situation changes little in the $4\times
8$ case, but has significantly changed for the $8\times 8$ case.
Obviously in the non-interacting case the dimensional cross-over
requires more than just a few legs.

While finite-size shell effects may explain the non-monotonic
variations we observed, we emphasize that the nature of the variation
of the pairing correlation to cross-over from one to two dimensions
was not revealed in our simulations. We note than even in the
apparently simpler classical Ising model the cross-over was slow.

\section{Concluding Remarks}


We presented an extensive QMC study of pairing correlations in the
attractive Hubbard model on chains, ladders, and squares. While still
computationally intensive, the QMC method we used is free of the
fermion sign problem that would prevent a similar study for the
repulsive Hubbard model. By using a method free of the sign problem,
we obtained results that apart from statistical errors are exact. In
several instances we re-enforced this exactness by reproducing
behavior predicted by exact analysis. For example, for chains we got
the correct asymptotic behavior of the on-site s-wave pairing
correlation function. For squares, we verified Zhang's relation
between the on-site and extended s-wave order parameters. Several of
our ambitious computational objectives were, however, limited by the
size of the systems and interactions strengths we could afford to
simulate. Still, we demonstrated the validity of the BCS approximation
in the weak coupling limit for a short-ranged interaction and
highlighted several important unresolved issues about finite-scaling of
fermion systems.

To a degree, the finite-size and shell effects exhibited by our
results are expected. Such effects have been reported before
\cite{furukama92,zhang95,guerrero98}. The
difficulty we had in trying to extrapolate beyond them to the
thermodynamic limit was unexpected. Part of the surprise follows from
QMC simulations of the attractive Hubbard model always having an
easier time revealing a signature of ODLRO than simulations of the
repulsive model. Years ago, in fact, ODLRO was reported for the
attractive model via a QMC simulation \cite{scalettar89}. How we
attempted to establish ODLRO in this paper however differs from how it
was attempted in that previous work. There, the uniform susceptibility
\begin{equation}
 \chi_s = \frac{1}{N}\sum_R P_s(R)
\end{equation}
was extrapolated to the thermodynamic limit. While this is an
appropriate quantity for very large system sizes, it is now
appreciated that for small systems, it is often dominated by the
values of $P_s(R)$ near $R=0$. These values merely measure local spin
and charge fluctuations. Directly examining the
long-distance behavior of $P_s(R)$, as done here, is a more
appropriate procedure.

Re-examining Fig.~3 of Ref.~\cite{scalettar89}, we notice the
following: The values of $\chi_s$ for small lattices were excluded from
the extrapolation. The values for the larger lattice sizes fluctuated
about the fit by amounts larger than the statistical error. We do not
doubt that a proper extrapolation will yield a non-negative value for
$\chi_s$. 
Noting that $\chi_s$ is roughly $\langle \Delta_s\rangle^2$, we remark
that their numerical values for $\langle \Delta_s\rangle$ are
consistent with ours.

From Fig.~3 of Ref.~\cite{scalettar89}, we also notice how smoothly the
uniform susceptibility for the charge-density wave extrapolated to the
thermodynamic limit. In the present work, we recall the smooth and
linear variation of the energy of the chains and the ladders with
size in contrast to the erratic variation for the squares.  The
finite-size effects influence different quantities and 
systems differently. While one expects these effects, where they
appear is a bit less predictable.

What is curious is the control the shell structure of the
non-interacting problem has on the behavior of the interacting
system. If one were to examine the variation of energy of the
two-dimensional Hubbard model as a function of electron density, one
would see shell locations defined by the non-interacting problem and
the energy in a shell to a very good approximation varying nearly
linearly with electron density
\cite{furukama92,zhang95,guerrero98}. The interactions do change the
slope (chemical potential) of this variation from the non-interaction
value; however, this nearly linear variation suggests the presence of
different sets of nearly degenerate ground states as the electron
density varies form shell to shell. One expects the interaction to
destroy the degeneracy in the shell of the non-interacting problem so
that shell effects in the interacting problem would be less than those
in the non-interacting problem. This expectation is not supported by
the data for all properties of the system.

One of our objectives was establishing how large must the system be
before we can say it is in a superconducting state. Over 40 years
ago, Anderson \cite{anderson58} asked how small can superconducting
metal particles be before they lose their superconducting
properties. He asserted the cut-off came when the gap near the Fermi
surface induced by the small size becomes larger than the
superconducting gap. There is evidence for the validity of this
condition from experimental studies of the superconducting properties
of metal clusters. What happens if the superconductor, as is the case
for the attractive Hubbard model, is gapless? On the other hand, is
the finite attractive model approximately a gaped superconductor? If
so, are our results suggesting the gaps are small?

What are the implications of our results for establishing ODLRO in the
repulsive Hubbard model? The behavior of the repulsive model is quite
different than the behavior of the attractive model. For any size of
the repulsive model studied so far, the difficulty is finding pairing
correlations larger than those of the non-interacting problem
\cite{zhang97}. In the thermodynamic limit the non-interacting problem
is not superconducting so whatever long-range correlations seen for
small systems must suppressed in larger systems. In the repulsive
model, d-wave pairing correlations are stronger than the s-wave pairing
correlations, and as the system size was increased, the magnitude of
the d-wave pairing correlations systematically vanished
\cite{zhang97,guerrero99}. What our results underscore is the care
needed to establish that any observed enhancement of the d-wave pairing
correlations is an intrinsic effect and not a finite-size effect.

We comment that all results reported here are for a single electron
filling. All the properties studied are a function of filling so at
some other fillings, it might be easier to extrapolate to the
thermodynamic limit and it might be possible that the BCS approximation
remains in quantitative agreement with the QMC simulations for larger
values of $U/t$. 
In fact, a few simulations for $\langle n\rangle = 0.875$ for
$4\times4$, $6\times6$, and $8\times8$ lattices showed only a  weak
variation in the results for the pairing correlation function.

While we
simulated systems at other fillings, these simulations were neither
extensive or systematic enough to establish conclusions other than
those now being reported. Also in building up the ladder to a square,
we took quite long ladders which restricted the number of legs we
could afford to simulate. We made the ladder long so the pairing
properties of the chain clearly matched the predications of the Bethe
{\em ansatz} solution of the infinite chain. It might be possible to
see the dimensional cross-over and extrapolate to the bulk
two-dimensional properties from shorter but wider ladders. For
example, in Fig.~12 we show $P_s(R)$ for a $16\times 8$ ladder at
$U/t=2$. The flatness suggests possible ODLRO consistent with the
results for the square geometry.

What are possible ways of handling the finite-size effects? One way,
suggested by Bormann et al. \cite{bormann91}, is adjusting $U/t$ for
different system sizes so the BCS approximation for $P_s(R)$ closely
approximates the one from the QMC simulations, subtracting this
approximation from the QMC results, and then extrapolating the
differences to the thermodynamic limit. After this, then one would add
to this result the BCS results for an infinite system.  (Actually,
Bormann et al. did this for $\chi_s$.) We did not try this procedure
because it is related to another one, scaling the vertex
correction. Here, one replaces averages like $\langle c^\dagger
c^\dagger c c\rangle$ with $\langle c^\dagger c^\dagger c c\rangle -
\langle c^\dagger c\rangle\langle c^\dagger c\rangle$ and studies the
size dependence of the remainder which is called the vertex
contribution \cite{white89}.  In spot checks, we found no
significantly different scaling behavior.

Another way to remove finite-size effects is the phase
averaging method suggested for exact diagonalization studies by Loh et
al. \cite{loh88} and recently adopted for QMC simulations by Ceperley
\cite{ceperley99}. In this procedure, one replaces the hopping
amplitude $t$ at the boundary by $te^{i\phi}$ and obtains various physical
quantities as a function of $\phi$. Then one averages these quantities
over $\phi$. Loh et al. give a justification and demonstration of
the method. In exact diagonalization studies, one just does a sequence
of diagonalizations for different values of $\phi$. In a Monte Carlo
method for computational efficiency it is necessary to treat $\phi$ as
another stochastic parameter and then let the random walk do the
averaging.  To do this one needs to change the QMC method. In the
applications contemplated by Ceperley, this means changing form the
fixed-node to the fixed-phase method \cite{ortiz93}. In our case we
are developing an analog of the the fixed-phase method, called the
constrained phase \cite{ortiz99}. Hopefully we will be able to report
results from this method soon.

Finally, we would like to emphasize that this paper has been dealing
exclusively with systems having a homogeneous long-range phase
coherence. There is a rapidly growing body of experimental evidence
suggesting that inhomogeneously textured (intrinsically nanoscale)
phases characterize the quantum state of high temperature
superconductors. It is possible that the superfluid density
characterizing that state is inhomogeneous. We believe that an
extension of the attractive Hubbard model including inhomogeneous terms
(mimicking stripes) could be the starting point to understand the
fundamental problem of inhomogeneous superfluids.

\acknowledgments
We thank D. Abrahams, G. A. Baker, Jr., E. Dagotto, E. Domany,
J. Eroles, J. L. Lebowitz, D. Pines, D. J. Scalapino, and
S. A. Trugman for helpful discussions and suggestions. We also thank
N. Kawakami, A. Moreo, and F. Marsiglio for providing us with some
data, and T. Momoi for some comments and for sending us copies of his
works. This work was supported by the U. S. Department of Energy which
also provided computational support at NERSC.
\newpage

\appendix
\section{BCS Equations}

In this Appendix we present the key equations used in our BCS
calculations. Some are well documented while others are not. For
completeness and convenience we give them here.

First to establish notation, we take (\ref{eq:hubbard}) as the Hamiltonian 
for a lattice of $N$ sites  and rewrite it in ${\bf k}$-space
\begin{equation}
H = \sum_{{\bf k},\sigma} \epsilon_{\bf k} n_{{\bf k},\sigma} -
\frac{U}{2 N} \sum_{{\bf k,k',q} \atop \sigma
=\uparrow,\downarrow}   c^{\dagger}_{{\bf k},\sigma}
c^{\dagger}_{{\bf k'},-\sigma} c^{\;}_{{\bf k'}+{\bf
q},-\sigma}c^{\;}_{{\bf k}-{\bf q},\sigma} \ ,
\end{equation}
using $\epsilon_{\bf k} = - 2 t (\cos k_x+\cos k_y)$ and assuming
periodic boundary conditions in all spatial directions. We used the BCS
wave function (\ref{eq:bcswfn}) with ${\bf k}$-independent relative phase
$\varphi$
\begin{equation}
|{\rm BCS}(\varphi)\rangle = \prod_{\bf k}(u_{\bf k}+v_{\bf k} e^{i\varphi} 
c^{\dagger}_{{\bf k},\uparrow} c^{\dagger}_{-{\bf k},\downarrow}) | 0 \rangle
\ .
\end{equation} 
The effective mean-field Hamiltonian, resulting from the neglect of
pairing fluctuation, is 
\begin{equation}
H_{\rm BCS} = \sum_{{\bf k},\sigma} \xi_{\bf k} n_{{\bf k},\sigma} -
\Delta \sum_{\bf k} (c^{\dagger}_{{\bf k},\uparrow} c^{\dagger}_{-{\bf
k},\downarrow} + c^{\;}_{-{\bf k},\downarrow} c^{\;}_{{\bf k},\uparrow}
) + \frac{U N_e^2}{4 N} \ ,
\end{equation}
where $N_e = \langle \hat N_e \rangle = 2 \sum_{\bf k} v_{\bf
k}^2$ is the average number of electrons and $\Delta =
\frac{U}{N} \sum_{\bf k}
\langle c^{\dagger}_{{\bf k},\uparrow}c^{\dagger}_{-{\bf
k},\downarrow}\rangle =\frac{U}{N} \sum_{\bf k} u_{\bf k} v_{\bf k}$
is the BCS gap,
and $\xi_{\bf k} = \epsilon_{\bf k} - \mu - U N_e/2N$.
The value of the energy was calculated from
\begin{equation}
\langle H_{\rm BCS}\rangle = \langle H - \mu \hat{N}_e \rangle = 2 \sum_{\bf
k} (\epsilon_{\bf k}- \mu) v_{\bf k}^2 - \frac{U}{N} \left [ \left (
\frac{N_e}{2} \right )^2 + \left ( \frac{N \Delta}{U} \right )^2 \right
]
\end{equation} 
and $\mu$ and $\Delta$ were determined by self-consistently solving 
\begin{eqnarray}
\sum_{\bf k} \Biggl[1-\frac{\xi_{\bf k}}{E_{\bf k}}\Biggr] &=& N_e \ , \\
\frac{U}{2 N} \sum_{\bf k} E_{\bf k}^{-1}=1 \ ,
\end{eqnarray}
where $ E_{\bf k} = \sqrt{\xi_{\bf k}^2 + \Delta^2}$.  With these
values of $\mu$ and $\Delta$ we determined $u_{\bf k}$, $v_{\bf k}$
from
\begin{eqnarray}
2 u_{\bf k} v_{\bf k} &=& \frac{\Delta}{E_{\bf k}}  \ , \\  
u_{\bf k}^2 - v_{\bf k}^2 &=& \frac{\xi_{\bf k}}{E_{\bf k}} \ .
\end{eqnarray}

A quantity central to this work is the s-wave superconducting pairing
correlation function $P_s ({\bf R}) = \langle \Delta^{\dagger}_{s}({\bf R})
\Delta_{s}(0) \rangle$. Within the BCS approximation it is
\begin{equation}
P_s ({\bf R})  =  \left(\frac{\Delta}{U}\right)^2 +
\frac{N_e}{2N}\delta_{{\bf R},0} - F({\bf R}) \ ,
\end{equation}
where the function $F({\bf R})$ 
\begin{equation}
F({\bf R}) = \frac{1}{N^2} \sum_{\bf k,k'} e^{i ({\bf k'} - {\bf k})\cdot {\bf R}} \
u_{\bf k'}^2 v_{\bf k}^2
\end{equation}
vanishes in the limit $|{\bf R}| \rightarrow \infty$. 

To establish the proportionality between the s-wave and extended s-wave
order parameters, we considered the Heisenberg equation of motion for
$\Delta_s(i) = c_{i,\downarrow} c_{i,\uparrow}$
\begin{equation}
-i \frac{\partial \Delta_s(i)}{\partial t} = \left [ H_{\rm BCS},
\Delta_s(i)\right ] \ .
\end{equation}
It is straightforward to show that 
\begin{equation}
\left [ H_{\rm BCS},\Delta_s(i)\right ] = (U N_e/N + 2 \mu) \Delta_s(i) +
t \  \Delta_{s^*}(i) - \Delta (n_{i,\uparrow}+n_{i,\downarrow} - 1)\ .
\end{equation}
Then using $\langle \left [ H_{\rm BCS},\Delta_s(i)\right ]\rangle=0$
for an equilibrium state, we find
\begin{equation}
\langle \Delta_{s^*}(i) \rangle = \frac{-U - 2 \mu}{t} \langle
\Delta_{s}(i) \rangle = \frac{-U - 2 \mu}{t} \left ( \frac{\Delta}{U}
\right ) \ ,
\label{extends}
\end{equation}
where the extended s-wave pair field operator operator is defined by
\begin{equation}
\Delta_{s^*}(i) = \sum_{{\delta}=\pm \hat{x}, \pm \hat{y}}
(c_{i,\downarrow} c_{i+ {\delta}, \uparrow} - c_{i,\uparrow} c_{i+
{\delta}, \downarrow})\ .
\end{equation}
This fundamental relation is formally the same as the one found by
Zhang \cite{zhang90} for the exact solution of the Hubbard model. We
note, however, that the chemical potential and expectation values in
(\ref{extends}) are BCS ones.

In the strong coupling regime ($U/t \rightarrow\infty$), 
the chemical potential $\mu = -\frac{U}{2}$ and the gap $\Delta =
\frac{1}{2} U\sqrt{1-(N_e/N-1)^2}$, therefore
\begin{equation}
 \Biggl(\frac{\Delta}{U}\Biggr)^2 \rightarrow 
  \frac{1}{4}\Biggl[1-\Biggl(\frac{N_e}{N}-1\Biggr)^2\biggr]
\end{equation}
and in that limit $\langle \Delta_{s^*}\rangle = 0$ and
\begin{equation}
 |{\rm BCS}(0)\rangle = 
  \prod_i \Biggl[\sqrt{1-\frac{N_e}{2N}}+\sqrt{\frac{N_e}{2N}} 
   c_{i\uparrow}^\dagger c_{i\downarrow}^\dagger\Biggr]|0\rangle \ .
\end{equation}

While not used in any of the results reported here, another relation
we derived and found useful in other contexts is a simple way to
project from the BCS wave function the component corresponding to a
fixed number of particles.  Although in the thermodynamic limit one can
ignore exact particle number conservation, for a finite system, one
often cannot because the particle number fluctuations $\langle
\hat{N}_e^2 \rangle - \langle\hat{N}_e \rangle^2= 4 \sum_{\bf k} u_{\bf
k}^2 v_{\bf k}^2$, inherent in the BCS wave function, can be large. To
project out the $N_e$-particle component, one uses
\begin{equation}
| \Psi_{N_e} \rangle = \frac{1}{2 \pi} \int_0^{2 \pi} \! \! d \varphi \
e^{-i \varphi N_e/2} \ | {\rm BCS}(\varphi) \rangle \ .
\end{equation}
By directly doing the implied integration over the phase, one can show
that the amplitude $\langle {\rm BCS}(0)|\Psi_{N_e}\rangle = \langle
\Psi_{N_e} |
\Psi_{N_e} \rangle =
\omega_{N_e}$ where $\sum_{N_e} \omega_{N_e}=1$. To find $\omega_{N_e}$ one can
use the recursion relation
\begin{equation}
N_p \ \omega_{N_p} = \omega_{N_p-1} \sum_{\bf k} \left ( \frac{v_{\bf
k}}{u_{\bf k}} \right )^2 + \sum_{j=1}^{N_p-1} (-1)^j \ 
\omega_{N_p-(j+1)} \sum_{\bf k} \left ( \frac{v_{\bf k}}{u_{\bf k}}
\right )^{2(j+1)} \ ,
\end{equation}
where $N_p=N_e/2$ is the number of pairs. We also note that $| {\rm
BCS}(0) \rangle = \sum_{N_e} | \Psi_{N_e} \rangle$.

\newpage

\section{Ising Ladders} 
In this Appendix we state the main equations necessary to study the
cross-over from one to two dimensions in a system of classical Ising
spins. We present analytical expressions from which various observables
are derivable.  In particular, we state those equations necessary to
study the large-distance behavior of the spin-spin correlation
function near and at the two-dimensional critical point.

Because of the extensive literature on the Ising model, we will
present only the results relevant for a $L \times M$ lattice, which
have not been explicitly documented to our knowledge. As close as we
could, we followed the notation and methodology in Ref. \cite{mccoy}.

The $M$-ladder Ising Hamiltonian in the absence of an external
magnetic field is:
\begin{equation}
H = - J_1 \sum_{i = -M_-}^{M_+} \ \sum_{j = 1- \bar{N}}^{\bar{N}} S_{i,j} 
S_{i,j+1} - J_2 \sum_{i = -M_-}^{M_+ - 1} \ \sum_{j = 1-
\bar{N}}^{\bar{N}}
S_{i,j} S_{i+1,j}
\end{equation}
where cylindrical boundary conditions ($S_{i,1-\bar{N}} = S_{i,1+\bar{N}}$)
are used, $S_{i,j}=\pm 1$, and 
\begin{equation}
\left \{
\begin{array}{rcl}
M_+ + M_- + 1 &=& M \; \;\mbox{(number of rows)} \\
2 \bar{N} &=& L \; \;\mbox{(number of columns)}
\end{array}
\right.
\end{equation}
Fig.~13 presents a schematic representation of the model Hamiltonian
$H$. The partition function ${\cal Z}_{M L}$ is:
\begin{equation}
{\cal Z}_{M L} = (2 \cosh{(\beta J_1)})^{M L} \ (\cosh{(\beta J_2)})^{(M-1) L} 
\ {\rm Pf}[\mathcal{A}] \ ,
\end{equation}
where the Pfaffian of the antisymmetric matrix $\mathcal{A}$ is 
\begin{equation}
{\rm Pf}^2 [\mbox{$\mathcal{A}$}] = \det \mbox{$\mathcal{A}$} = 
\prod_{\theta} \left\{ \left | 1 + z_1 e^{i \theta} \right |^{2 M}  
\lambda^M [ v^2 + \alpha^{-2 M} \bar{v}^2 ]  \right\} \ ,
\end{equation}
with $\theta = \pi (2n -1)/L$ ($n=1,2,\cdots,L$), $z_{1(2)} = 
\tanh{(\beta J_{1(2)})}$, 
\begin{equation}
\lambda = \frac{z_2 (1 - z_1^2) \alpha}{\left | 1 + z_1 e^{i
\theta} \right|^{2}} \;\;\;, \;\;\; \bar{\lambda} = 
\frac{z_2 (1 - z_1^2)}{\left | 1 + z_1 e^{i
\theta} \right|^{2} \alpha}
\end{equation}
\begin{eqnarray}
\alpha &=& \frac{1}{2 z_2 (1 - z_1^2)} \left \{ (1 + z_1^2)(1 + z_2^2) -
z_1 (1 - z_2^2) \left [ e^{i \theta} +  e^{-i \theta} \right ] \right .
\nonumber \\
&&  (1 - z_2^2) \sqrt{(1 - \alpha_1 e^{i \theta})(1 - \alpha_1
e^{-i \theta}) \left (1 - \frac{e^{i \theta}}{\alpha_2} \right )
\left (1 - \frac{e^{-i \theta}}{\alpha_2} \right ) } \Biggr \}
\end{eqnarray}
\begin{equation}
\alpha_1 = \frac{z_1 (1 - |z_2|)}{1 + |z_2|} \;\;\;, \;\;\;
\alpha_2 = \frac{z_1^{-1} (1 - |z_2|)}{1 + |z_2|} 
\end{equation}

\begin{equation}
v^2 = \frac{1}{2} \left ( 1 + 
\frac{z_2^2 - (4 z_1^2 \sin^2{(\theta)}+ (1-z_1^2)^2)/ \left | 1 + z_1 e^{i
\theta} \right |^4}{\bar{\lambda} - \lambda} \right ) \;\;\;, \;\;\;
\bar{v}^2 = 1 - v^2 
\end{equation}

We were mainly interested in studying the spin-spin correlation
function along the central row (as indicated in Fig. 13).  This function
is given by
\begin{equation}
\langle S_{0,0} S_{0,n} \rangle^2 = (1-z_1^2)^{2 n} \det 
\left[\begin{array}{ccc}
    T_{R R} & T_{R L} \\
    T_{L R} & T_{L L}
\end{array}\right]
\end{equation}
where the $n \times n$ matrices $T_{I J}$ are
\begin{equation} 
T_{R R} = 
\left[\begin{array}{ccc}
    0 & \cdots & A^{-1}(0;n-1)_{R R} \\
    A^{-1}(1;0)_{R R} & \cdots & A^{-1}(1;n-1)_{R R} \\
    \vdots & & \vdots \\
    A^{-1}(n-1;0)_{R R} & \cdots & 0
\end{array}\right]
\end{equation}
\begin{equation} 
T_{R L} = 
\left[\begin{array}{ccc}
    A^{-1}(0;1)_{R L} - c & \cdots & A^{-1}(0;n)_{R L} \\
    A^{-1}(1;1)_{R L} & \cdots & A^{-1}(1;n)_{R L} \\
    \vdots & & \vdots \\
    A^{-1}(n-1;1)_{R L} & \cdots & A^{-1}(n-1;n)_{R L} - c
\end{array}\right]
\end{equation}
\begin{equation} 
T_{L R} = 
\left[\begin{array}{ccc}
    A^{-1}(1;0)_{L R} + c & \cdots & A^{-1}(1;n-1)_{L R} \\
    A^{-1}(2;0)_{L R} & \cdots & A^{-1}(2;n-1)_{L R} \\
    \vdots & & \vdots \\
    A^{-1}(n;0)_{L R} & \cdots & A^{-1}(n;n-1)_{L R} + c
\end{array}\right]
\end{equation}
\begin{equation} 
T_{L L} = 
\left[\begin{array}{ccc}
    0 & \cdots & A^{-1}(1;n)_{L L} \\
    A^{-1}(2;1)_{L L} & \cdots & A^{-1}(2;n)_{L L} \\
    \vdots & & \vdots \\
    A^{-1}(n;1)_{L L} & \cdots & 0
\end{array}\right]
\end{equation}
with $c = (z_1^{-1} - z_1)^{-1}$, $A^{-1}(k;k')_{I J} = 
\frac{1}{L} \sum_{\theta} e^{i \theta (k - k')} \left [ 
B^{-1}(\theta) \right]_{I J}$ and 
\begin{eqnarray}
\left [ B^{-1}(\theta) \right]_{R R} = - \left [ B^{-1}(\theta) \right]_{L L} 
&=& \frac{1}{\left | 1 + z_1 e^{i \theta} \right |^2} \left \{ 
\left [ b_{22}^{-1} \right ]_{UU} + \left
[ b_{22}^{-1} \right ]_{DD} \right \} \\
\left [ B^{-1}(\theta) \right]_{R L} = - \left [ B^{-1}(\theta) 
\right]_{L R}^{*} 
&=& \frac{-1}{(1 + z_1 e^{-i \theta})} \left \{ 1 - 
\frac{1}{(1 + z_1 e^{-i \theta})} \left (
\left [ b_{22}^{-1} \right ]_{UU} - 
\left [ b_{22}^{-1} \right ]_{DD} -
\left [ b_{22}^{-1} \right ]_{DU} +
\left [ b_{22}^{-1} \right ]_{UD} \right ) \right \} 
\end{eqnarray}

\begin{eqnarray}
\left [ b_{22}^{-1} \right ]_{UU} &=& i \frac{v \bar{v}}{z_2} \left (1 - 
\alpha^{-2(M_-+1)} \right ) \frac{v^2 + \bar{v}^2 \alpha^{-2 
M_+}}{v^2 + \bar{v}^2 \alpha^{-2 M}} \\
\left [ b_{22}^{-1} \right ]_{DD} &=& -i \frac{v \bar{v}}{z_2} \left (1 - 
\alpha^{-2(M_++1)} \right ) \frac{v^2 + \bar{v}^2 \alpha^{-2 
M_-}}{v^2 + \bar{v}^2 \alpha^{-2 M}} \\
\left [ b_{22}^{-1} \right ]_{UD} &=& \frac{\left (v^2 + \bar{v}^2 \alpha^{-2 
M_+} \right ) \left (v^2 + \bar{v}^2 \alpha^{-2 M_-} \right )}{z_2 \alpha 
\left (v^2 + \bar{v}^2 \alpha^{-2 M_-} \right )}  \\
\left [ b_{22}^{-1} \right ]_{DU} &=& -\left [ b_{22}^{-1} \right ]_{UD}
\end{eqnarray}

If one were interested in the strip geometry ($L \rightarrow \infty$),
then $T_{R R} = T_{L L} = 0$ and $\langle S_{0,0} S_{0,n} \rangle$
could be written in terms of an $n \times n$ Toeplitz determinant.
\newpage


%
%

\begin{figure}
\caption{The on-site s-wave pairing correlation function $P_s(|R|)$ as a
function of the distance $|R|$ between pairs for a $U/t=2$
chain of length $L=50$. (a) Comparison of the results of the BCS
approximation with a QMC
calculation that used (\ref{eq:standard}) and a DMRG calculation that
used (\ref{eq:average}). (b) Comparison of the DMRG results for the
two different ways, (\ref{eq:center}) and (\ref{eq:average}), of
estimating the correlation function.}
\label{fig2}
\end{figure}

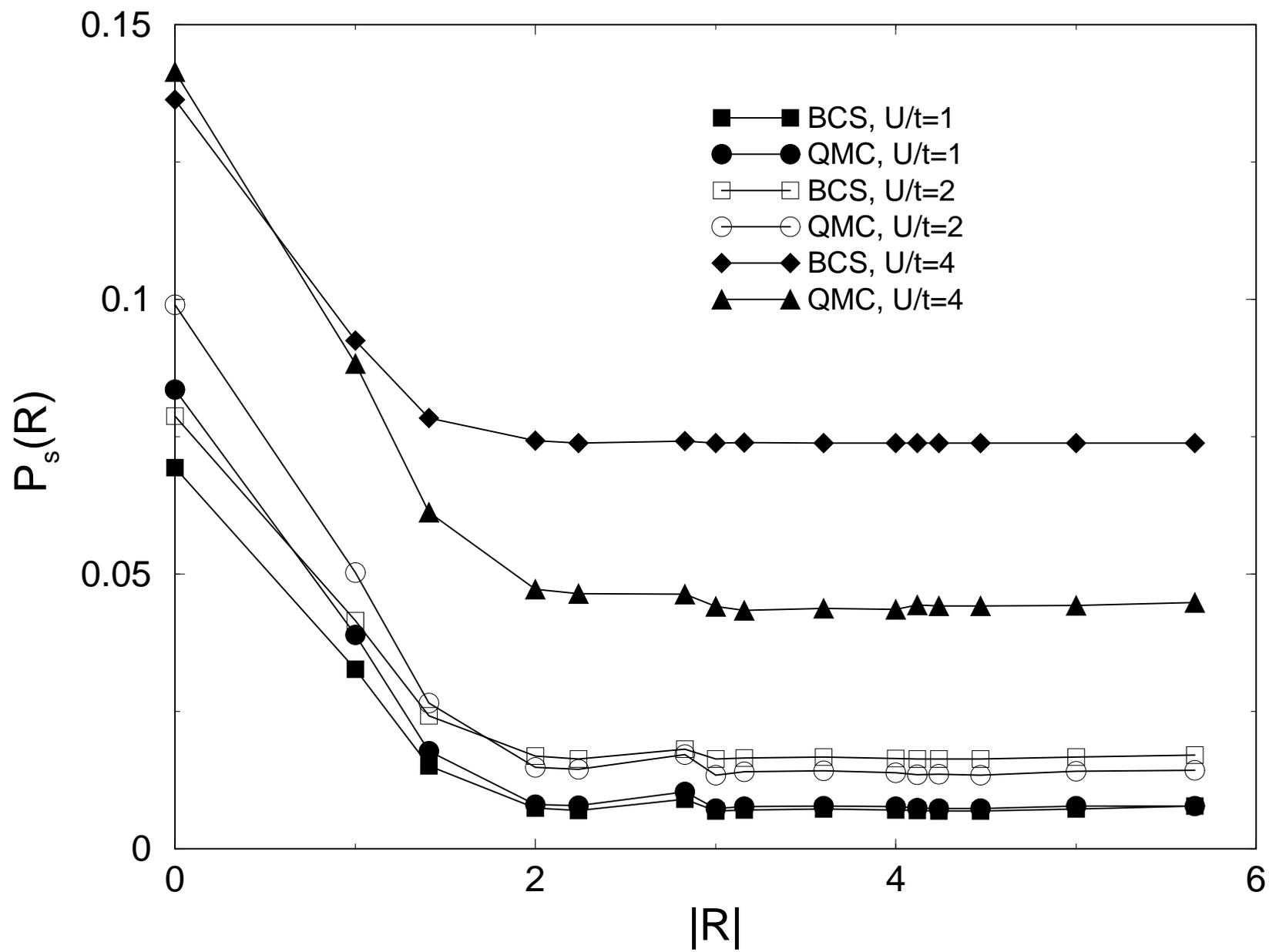
\begin{figure}
\caption{The QMC and BCS on-site s-wave pairing correlation function
$P_s(R)$ as a function of the distance $|R|$ between pairs for a
$8\times 8$ lattice and $U/t=1$, $U/t=2$, and $U/t=4$.}
\label{fig3}
\end{figure}

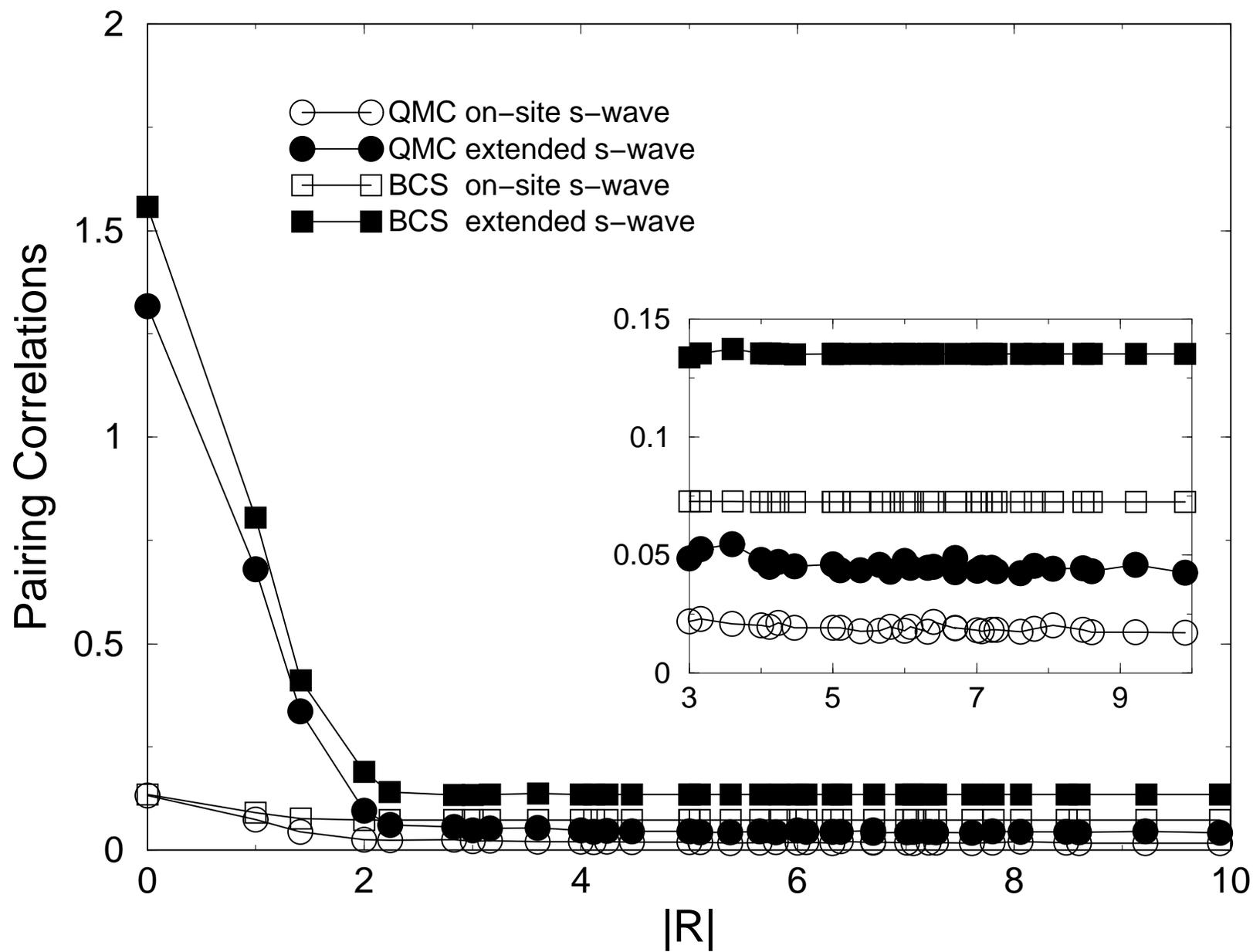
\begin{figure}
\caption{The QMC and BCS on-site and extended s-wave pairing
correlation functions $P_s(R)$ as a function of the distance $|R|$
between pairs for $U/t=2$ and a $14\times 14$ lattice. The inset shows the
large distance behavior of these same functions.}
\label{fig4}
\end{figure}

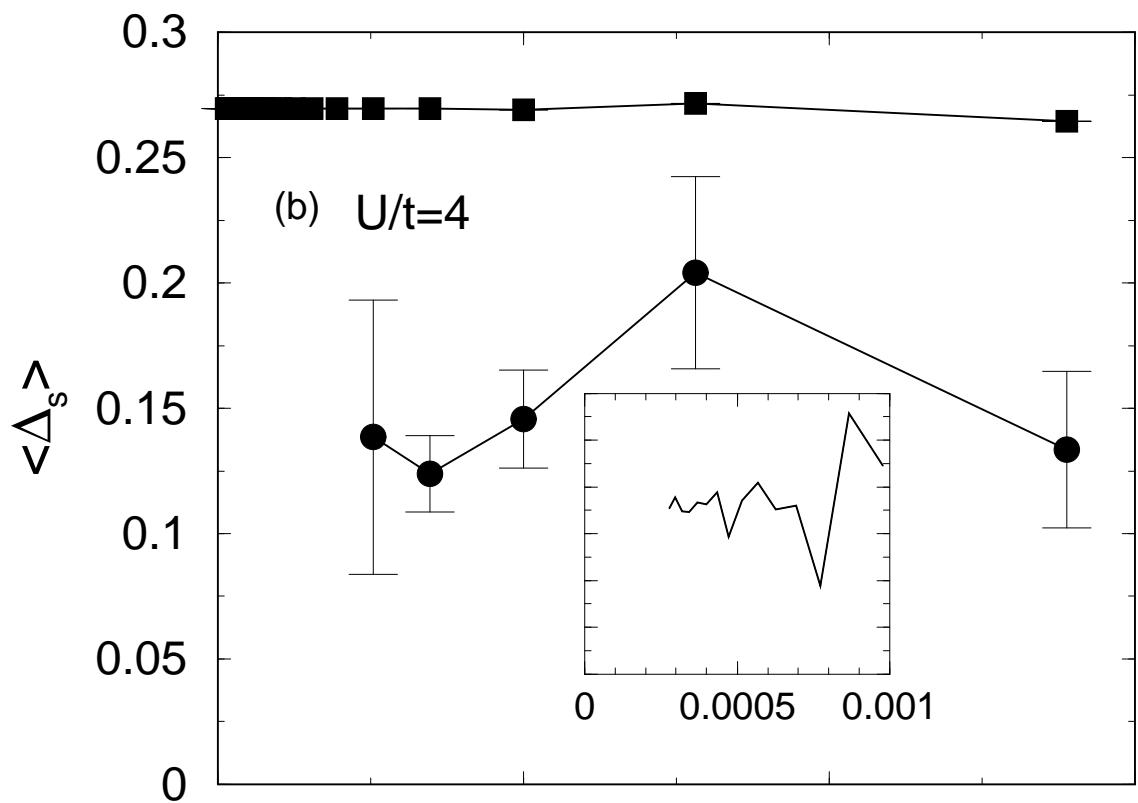
\begin{figure}
\caption{The expectation value of the on-site s-wave
order parameter $\langle \Delta_s\rangle$ as a function of the
reciprocal of the size $N$ of a square lattice. In both (a) and (b)
the QMC results are compared to those of the BCS approximation. In
(a), $U/t=2$. In (b), $U/t=4$ and the inset shows the behavior of the BCS
predictions for small values of the reciprocal of the lattice size.}
\label{fig5}
\end{figure}

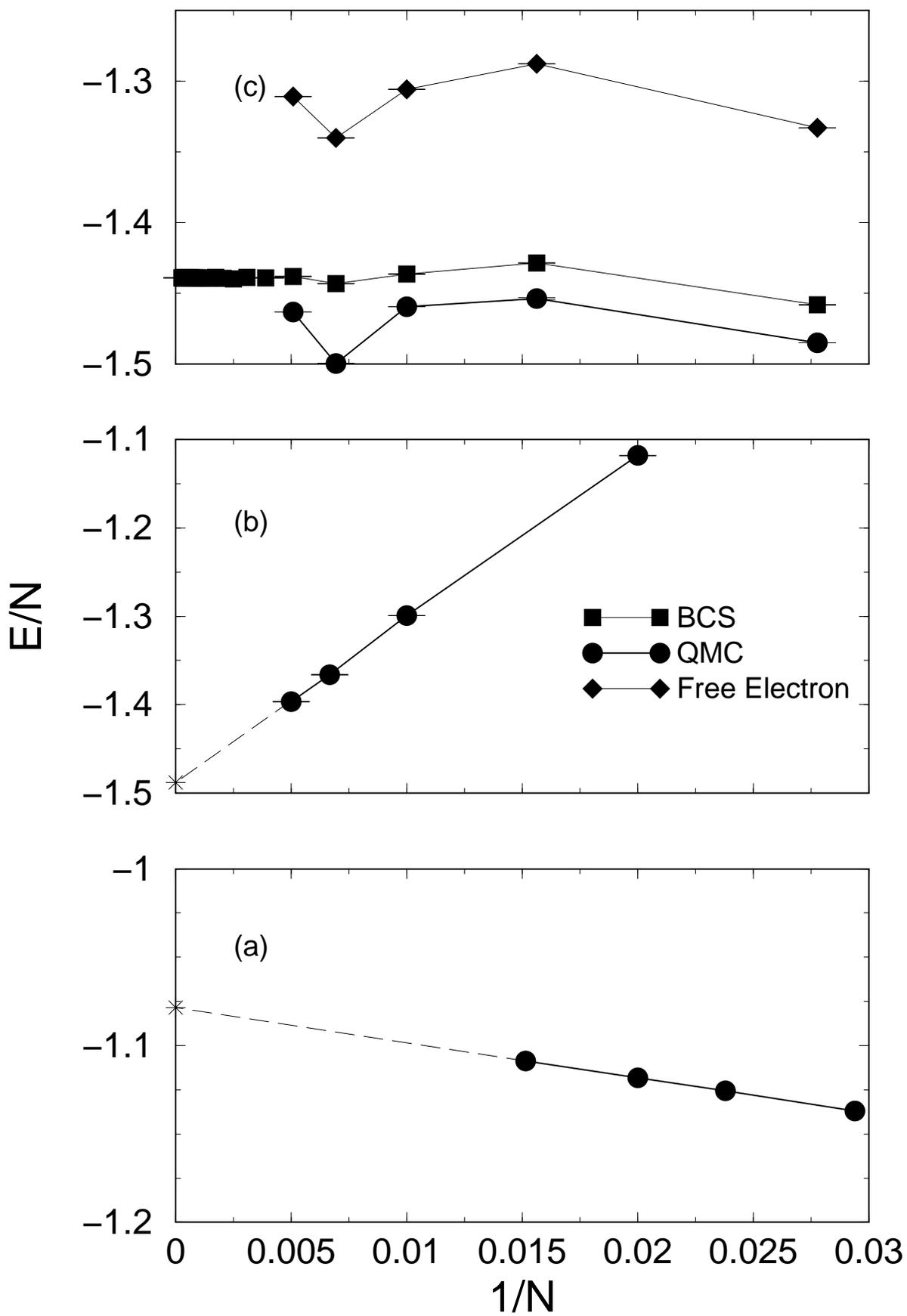
\begin{figure}
\caption{The QMC energy per site $E/N$ as a function of the reciprocal of
the number $N$ of lattice sites for $U/t=2$ systems. (a) Chains
($N=L$), (b) ladders ($N=L\times M$), and (c) squares ($N=L\times
L$). For squares the free-electron and BCS results are also shown. For
chains and ladders, the QMC curves smoothly and linearly extrapolate
to the thermodynamic limit indicated by star-burst symbol at
$1/N=0$. }
\label{fig6}
\end{figure}

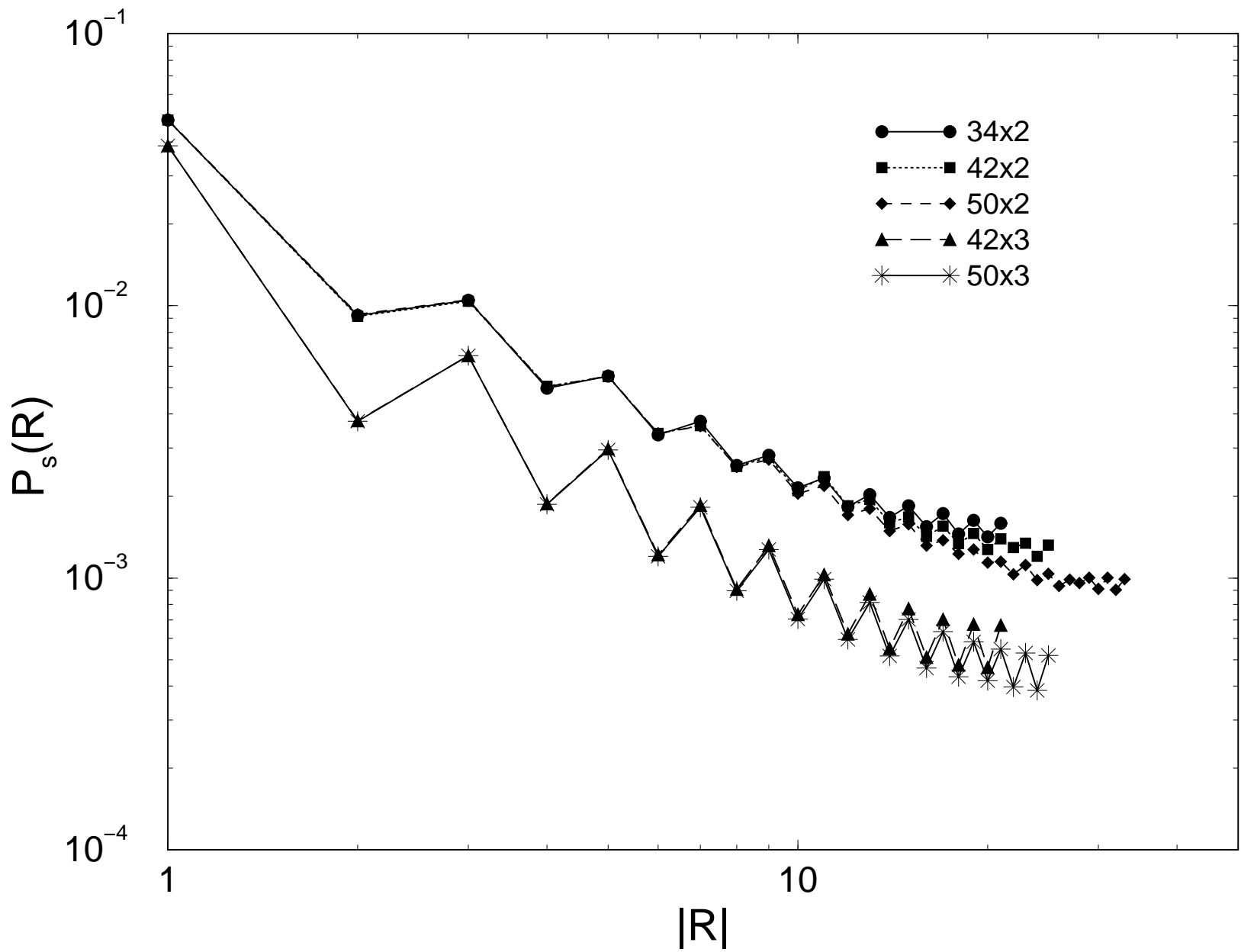
\begin{figure}
\caption{The QMC on-site s-wave pairing correlation function $P_s(R)$ as a
function of the distance $|R|$ between pairs for $U/t=2$, and 2 and 3-legged
ladders of different lengths.}
\label{fig7}
\end{figure}

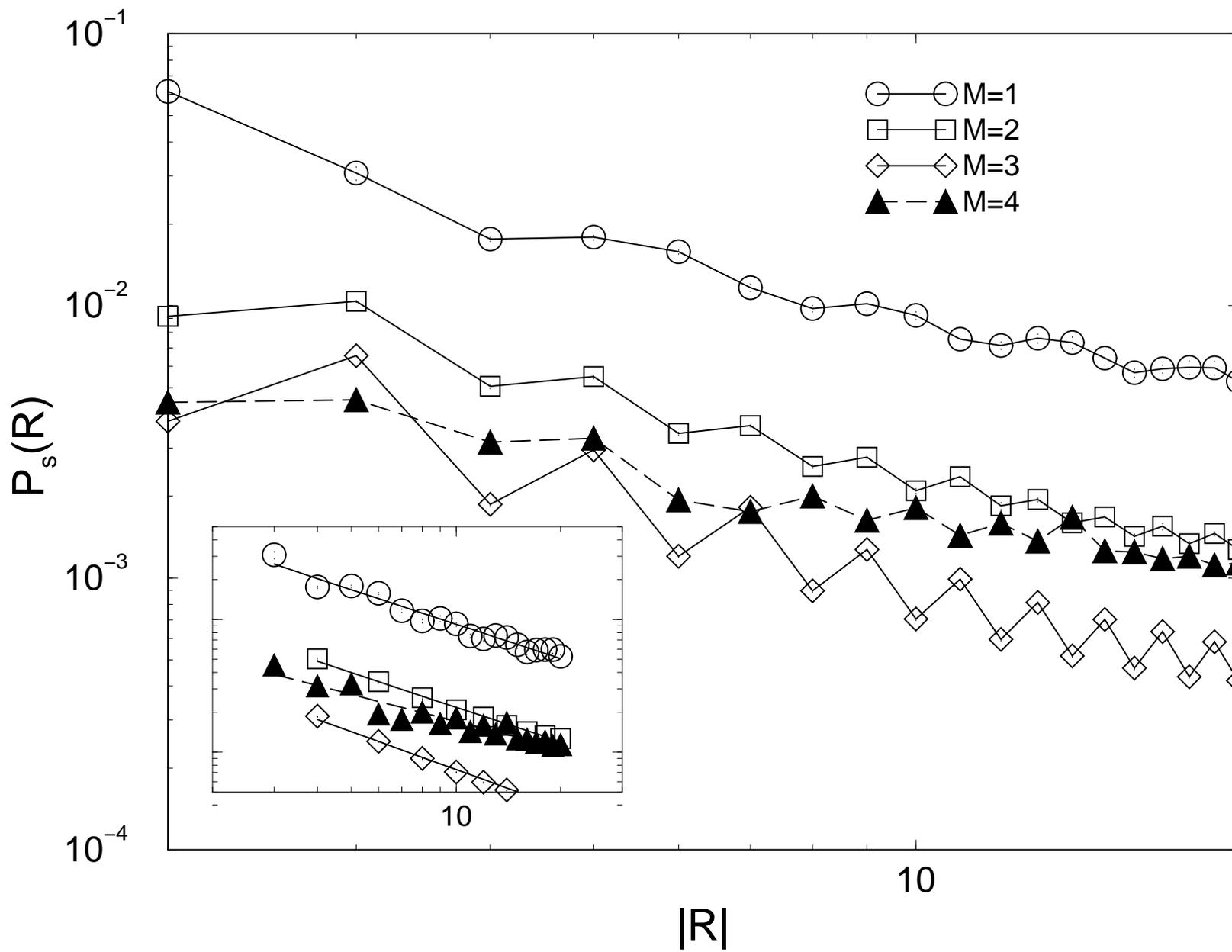
\begin{figure}
\caption{The QMC on-site s-wave pairing correlation function $P_s(R)$ as a
function of the distance $|R|$ between pairs for a $U/t=2$ ladder of
length 50 as a function of the number of legs $M$. The inset shows
the long-range behavior of these correlations fitted to the
inverse-power law function (\ref{eq:power})}
\label{fig8}
\end{figure}

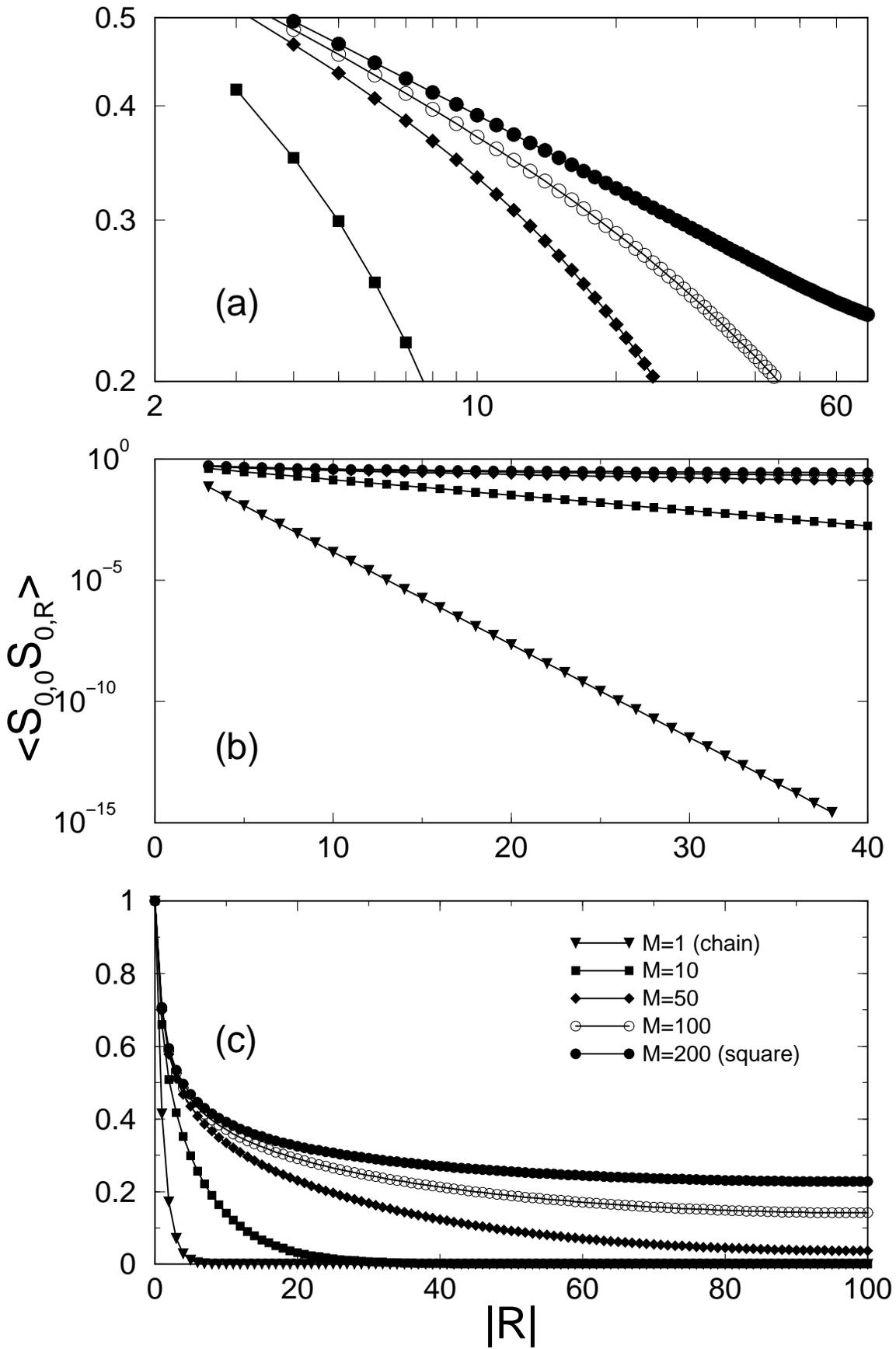
\begin{figure}
\caption{The spin-spin correlation function for an Ising 
ladder of length $L=$200 as a function of the distance $|R|$ between spins
and of the number of legs $M$. The temperature is at the bulk critical
value, and the distance $|R|$ is along a central leg in the
$x$-direction. Cylindrical boundary conditions were used.}
\label{fig9}
\end{figure}

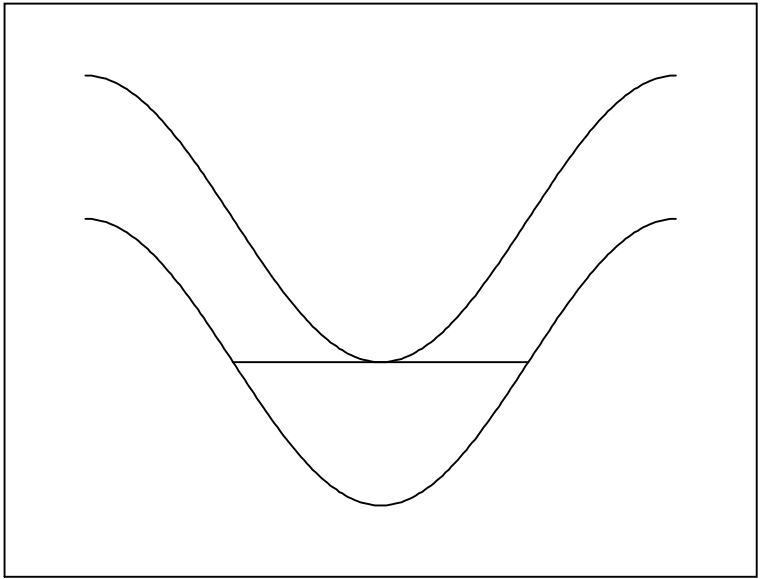
\begin{figure}
\caption{Band structure of the non-interacting problem on 1, 2, 3, and
4 legged ladders. Cylindrical boundary conditions are applied. The long
length $L$ is infinite, and the horizontal line represents the Fermi energy
for a quarter-filled systems.}
\label{fig10}
\end{figure}

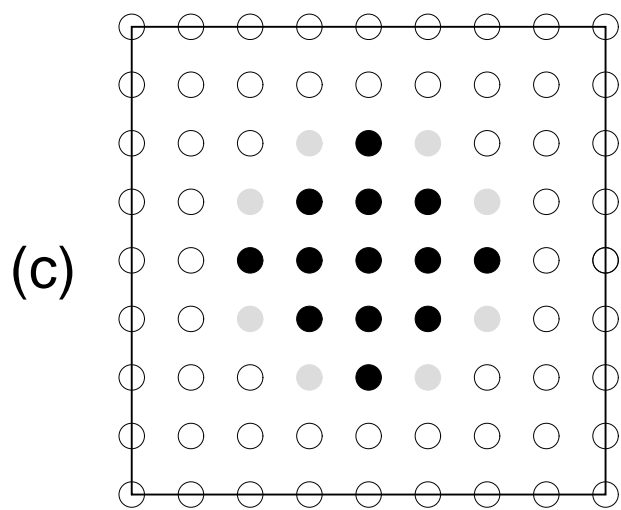
\begin{figure}
\caption{The allowed {\bf k}-states for a $2\times 8$, $4\times 8$, and
$8\times 8$ non-interacting problem with periodic boundary
conditions. Illustrated are which {\bf k}-states are occupied for
quarter-filling of an equal number of up and down spin electrons. A
black dot denotes double occupancy, an open circle represents no
occupancy, and a grey dot represents the degenerate state comprising
the open shell.}
\label{fig11}
\end{figure}

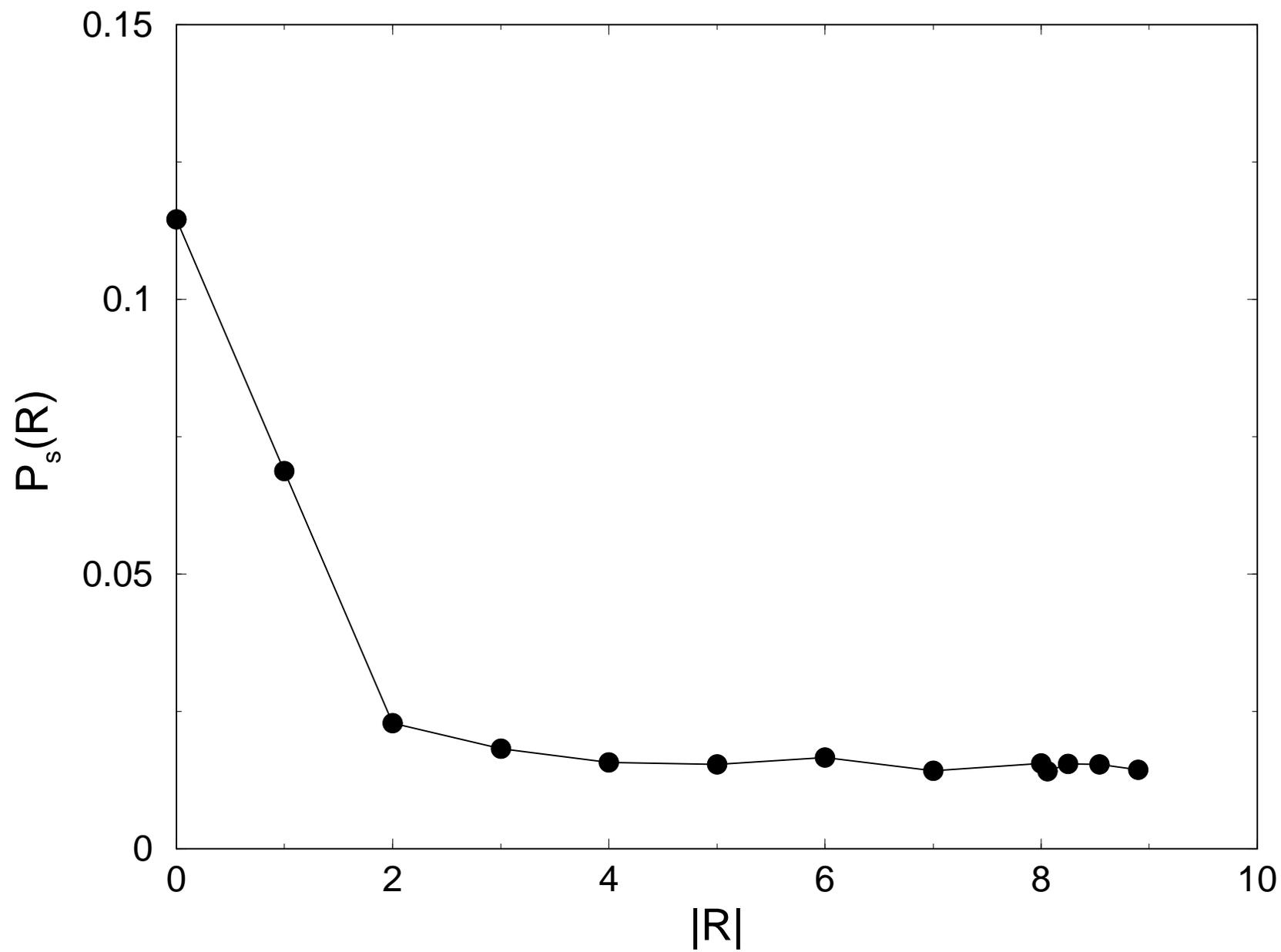
\begin{figure}
\caption{The QMC on-site s-wave pairing correlation function $P_s(R)$ as a
function of the distance $|R|$ between pairs for a rectangular quarter-filled
$U/t=2$ lattice of size $16\times 8$.}
\label{fig12}
\end{figure}

\begin{figure}
\caption{The cylindrical geometry used for the Ising model ladder with
$M$ legs.}
\label{fig13}
\end{figure}

%
%
\begin{table}
\caption{Summary of relevant parameters of the QMC simulations. Shown
are the systems sizes, the
electron fillings $\langle n \rangle$, and the
exponent $\beta$ of the power law decay for $U/t=2$. For 2 and 3-legged 
ladders, where there is a 
double entry, for example, $50 \times 2$, the top entry gives the
value of $\beta$ for the top envelope, while the bottom entry is for
the bottom envelope. In one dimension the fillings are always
$\langle n \rangle = 1/2 + 1/L$}
\label{table1}
\begin{tabular}{rdd}
System Size & $\langle n \rangle$ & $\beta$ \\ \hline
$34 \times  1$ & 0.53 & 0.88(9) \\
$42 \times  1$ & 0.52 & 0.82(6) \\
$50 \times  1$ & 0.52 & 0.86(4) \\
$66 \times  1$ & 0.52 & 0.79(3) \\ \hline
$50 \times  1$ & 0.52 & 0.86(4) \\
$50 \times  2$ & 0.50 & 1.07(3) \\
               &      & 0.87(2) \\
$50 \times  3$ & 0.51 & 1.32(4) \\
               &      & 0.95(3) \\
$50 \times  4$ & 0.51 & 0.68(5) \\ \hline
$ 6 \times  6$ & 0.5  &         \\
$ 8 \times  8$ & 0.5  &         \\
$10 \times 10$ & 0.5  &         \\
$12 \times 12$ & 0.51 &         \\
$14 \times 14$ & 0.5  &         \\ \hline
$16 \times  8$ & 0.45 &         \\ 
\end{tabular}
\end{table}

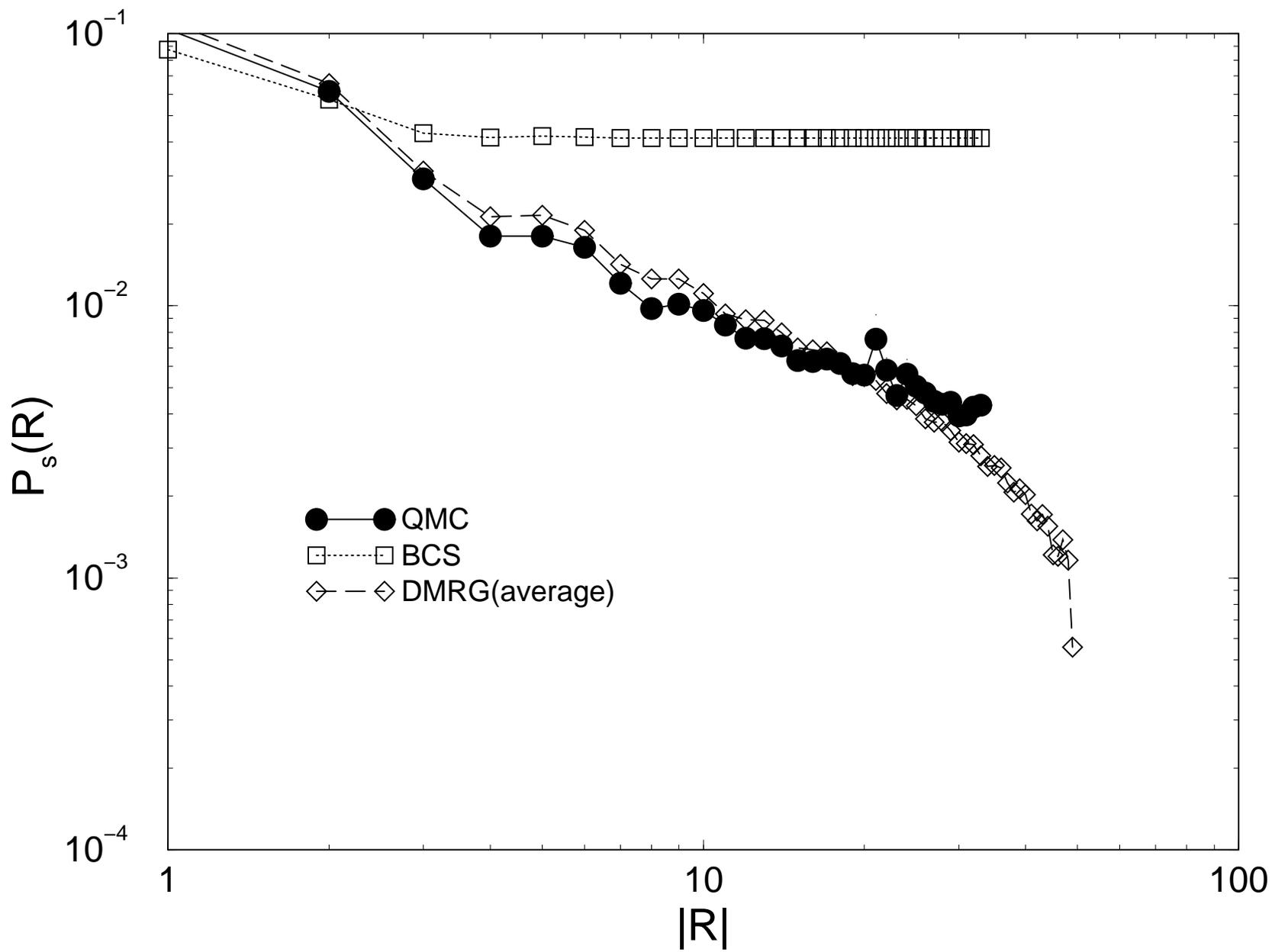
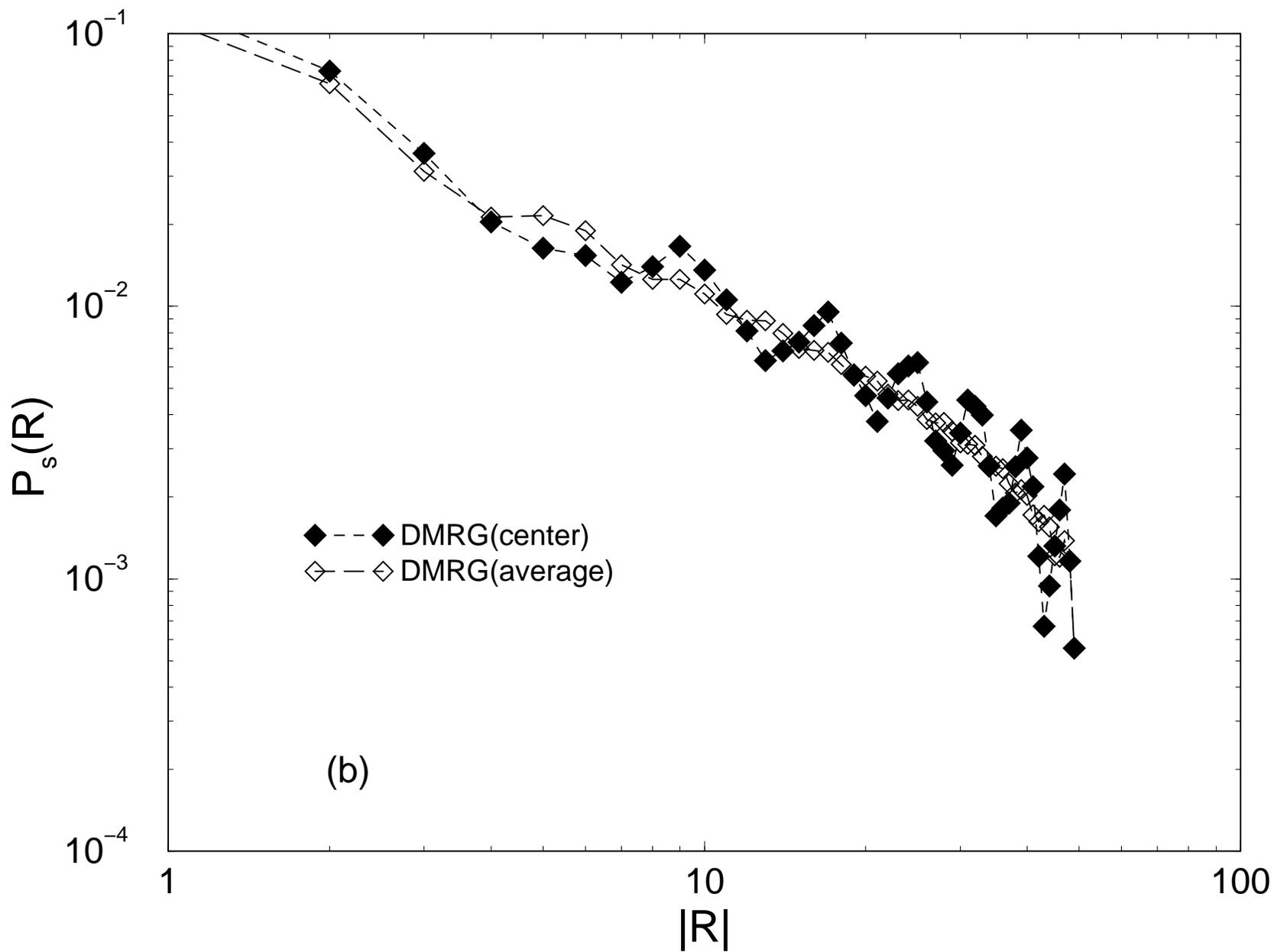
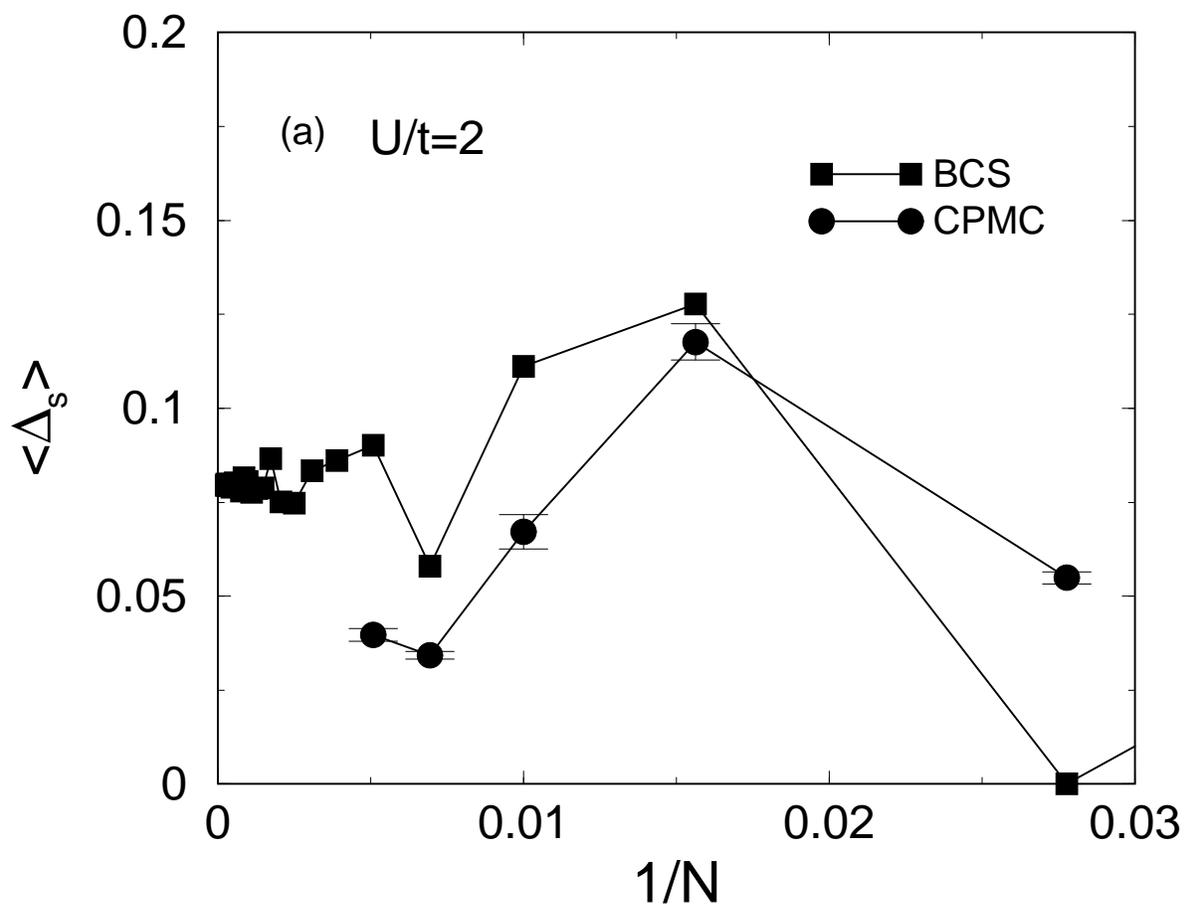
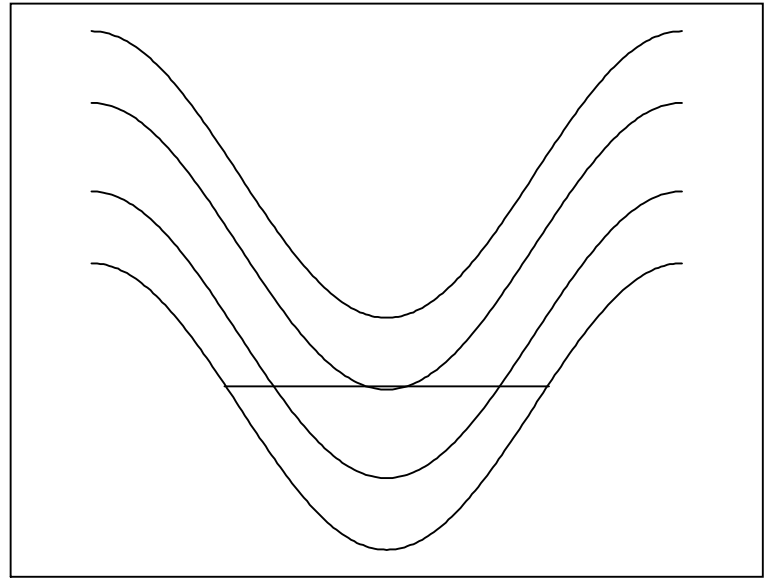
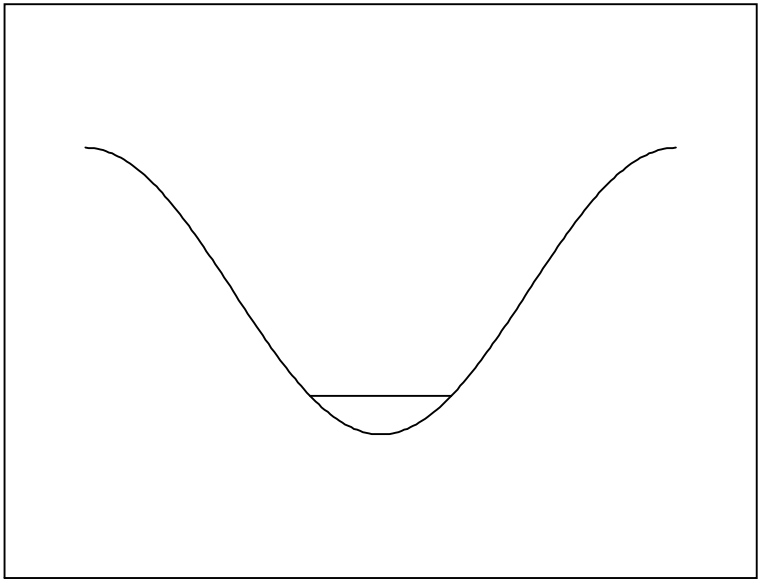
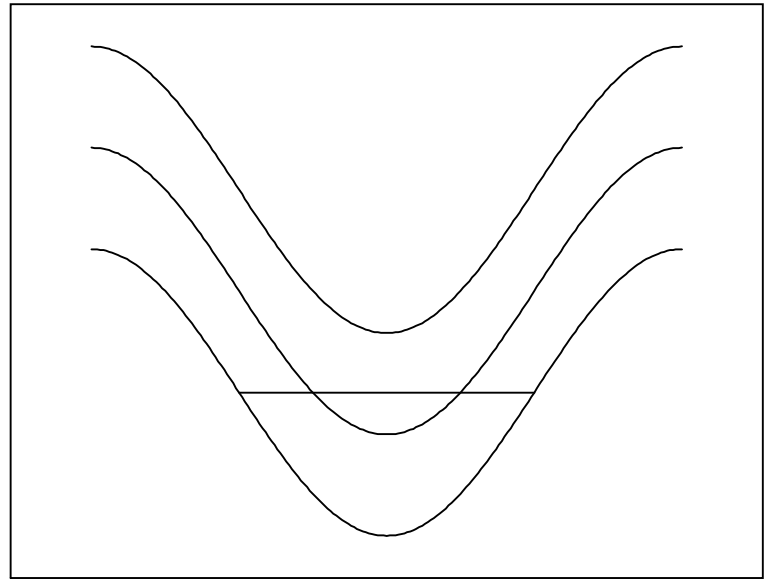
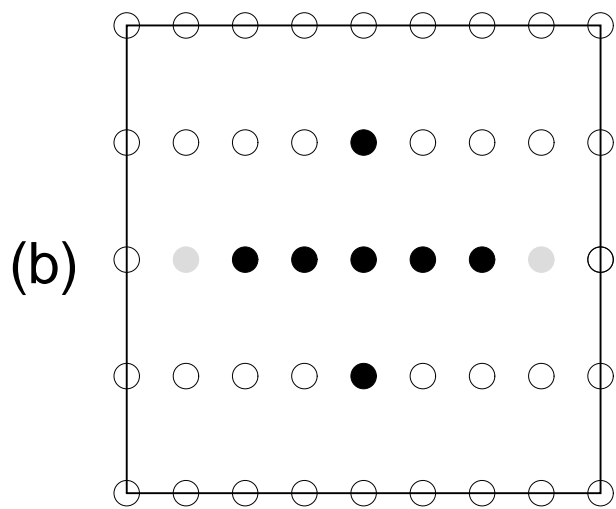
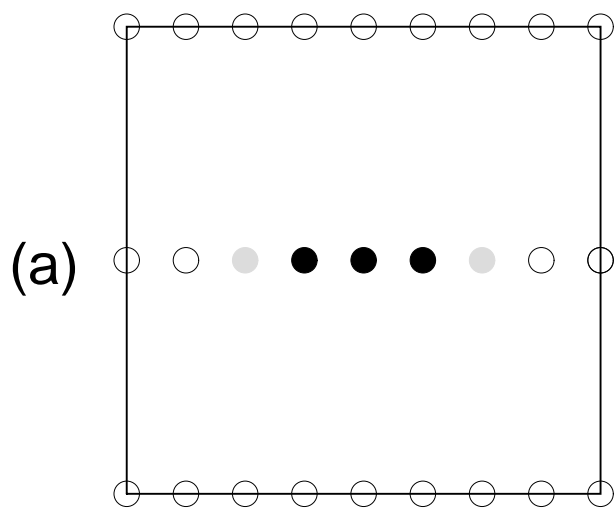
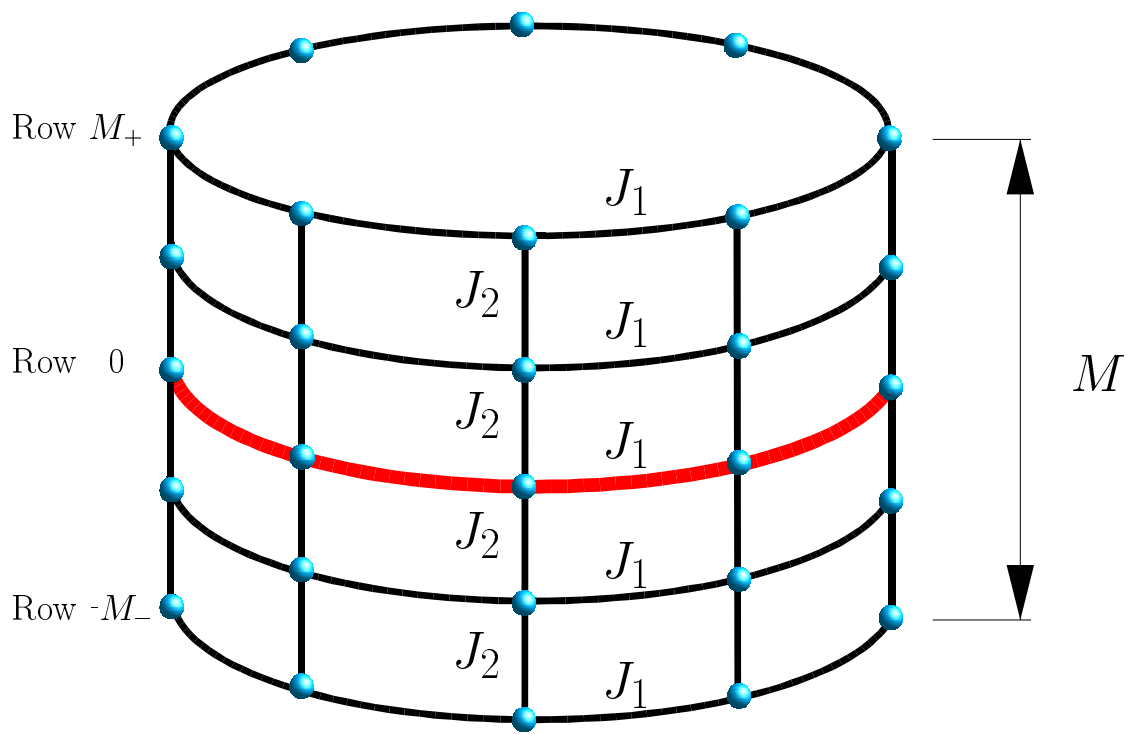
\end{document}